\documentclass[a4paper,
10pt
]{article}

\usepackage[english]{babel}

\usepackage[a4paper,left=3cm,right=3cm,top=2.5cm,bottom=2.5cm]{geometry}

\usepackage{amsfonts} 
\usepackage{hyperref}
\usepackage{bm}
\usepackage{authblk}
\usepackage{siunitx}
\usepackage{graphicx}
\usepackage[english]{babel}
\usepackage{amsmath}
\usepackage{xcolor}
\usepackage{subcaption}
\usepackage{float}
\usepackage{multirow}
\newcommand{\fig}[1]{Fig.~\ref{#1}}{\color{blue}}
\newcommand{\tab}[1]{Tab.~\ref{#1}}
\usepackage{comment}
\usepackage{stackengine,graphicx}
\usepackage{tikz}
\usepackage{tabularx}
\usepackage[linesnumbered,ruled,vlined]{algorithm2e}
\usepackage[draft,inline,marginclue]{fixme}
\FXRegisterAuthor{gs}{ags}{\color{red}GS}

\title{A Deep-Learning Enhanced Gappy Proper Orthogonal Decomposition Method for Conjugate Heat Transfer Problem}

\author[1]{Arash Hajisharifi\thanks{These author(s) contributed equally to this work.}\thanks{Emails: ahajisha@sissa.it, rhalder@sissa.it, mgirfogl@sissa.it, giovanni.stabile@santannapisa.it, grozza@sissa.it}}
\author[1]{Rahul Halder\thanks{These author(s) contributed equally to this work.}}
\author[1]{Michele Girfoglio}
\author[2]{Giovanni Stabile}
\author[1]{Gianluigi Rozza}

\affil[1]{Mathematics Area, mathLab, SISSA, via Bonomea 265, I-34136 Trieste, Italy}
\affil[2]{Biorobotics Institute, Sant'Anna School of Advanced Studies, Piazza Martiri della Libertà, 33, 56127 Pisa, Italy}
\date{}

\begin{document}
\maketitle
% \listoffixmes

%\maketitle

\begin{center}
\bf ABSTRACT
\end{center}

The current study aims to develop a non-intrusive Reduced Order Model (ROM) to reconstruct the full temperature field for a large-scale industrial application based on both numerical and experimental datasets. The proposed approach is validated against a domestic refrigerator. At the full order level, air circulation and heat transfer in fluid and between fluid and surrounding solids in the fridge were numerically studied using the Conjugated Heat Transfer (CHT) method to explore both the natural and forced convection-based fridge model followed by a parametric study-based on the ambient temperature, fridge fan velocity, and evaporator temperature. %Considering the computational cost of the full order problem, a non-intrusive reduced order model based on \textcolor{blue}{proper orthogonal decomposition (POD) with the interpolation method such as Radial basis function (POD-RBF) and with a regression approach such as Artificial Neural Network (POD-ANN)} are first considered to obtain the temperature field at specific parametric locations where the training dataset is purely based on numerical computation. Furthermore, a Gappy Proper Orthogonal Decomposition (GPOD) based approach is employed to reconstruct the full temperature field with data available at a few sparse sensor locations which can potentially arise from both numerical computation or experiments.
The main novelty of the current work is the introduction of a stable Artificial Neural Network (ANN) enhanced Gappy Proper Orthogonal Decomposition (GPOD) method which shows better performance than the conventional GPOD approach in such large-scale industrial applications. The full-order model is validated with the experimental results and the prediction accuracy of the surrogate model associated with different reduced-order approaches is compared with the benchmark numerical results or high-fidelity results. In our current work, we show that a prediction error of $1$ ($^o C$) and computational speed-up of $5e3$ is achieved even at a very sparse training dataset using the proposed deep-learning enhanced GPOD approach. 

\maketitle

\section{Introduction}

%\GS{this part should we written really carefully, highlighting differences wrt existing works, citing properly literature and trying to sell the article in the best possible way}

Significant research has been carried out on ROM in the last few decades. Its application ranges in several industrial practices such as aerospace \cite{ripepi2018reduced, salavatidezfouli2025modal}, pharmaceuticals \cite{hajisharifi2023non, hajisharifi2024lstm} , marine engineering \cite{demo2018shape, demo2018efficient} and even home appliances industries \cite {cole2014reduced, bayer2013cfd, xu2023co}.

Domestic refrigerator system is one key application of conjugate heat transfer problems driven by forced and natural convection. Maintaining the prescribed limit of the temperature fields inside different compartments is one of the key criteria of the domestic refrigerator design to ensure energy efficiency and freshness of food.

Several research works on heat transfer in an enclosed chamber such as work by Markatos et al.\cite{markatos1984laminar} provide the necessary numerical tools required for the development of the full computational fluid dynamics-based domestic fridge model governed by natural convection. A full-scale domestic fridge with natural convection was modeled by Laguerre and Flick \cite{laguerre2004heat}. Numerical simulation considering the radiation effect was carried out in their later work \cite{laguerre2007numerical}. $\text{S{\"o}ylemez}$ et al. \cite{soylemez2021cfd} considered forced convection arising due to the fan velocity inside a domestic refrigerator. However, in their numerical model, only the fluid volume was considered neglecting the heat transfer in the solid region of the refrigerator. Zhang and Lian \cite{zhang2014conjugate} considered the conjugate heat transfer (CHT) algorithm to account for heat transfer in both the fluid and solid regions of the domestic refrigerator model. However, there are several research works carried out to numerically model a domestic refrigerator, there is limited work considering all the relevant physics associated with the system such as conjugate heat transfer, natural and forced convection, radiation heat transfer and turbulence.   

The full-order numerical model associated with the domestic refrigerator is numerically expensive especially when there are repetitive tasks involved such as in the design step, process optimization, and development of control
algorithm. The reduced order model approach facilitates the exploration of parametric space while carrying out those repetitive tasks in industrial practice. In general, the reduced order model consists of two steps - one is the offline phase where the high-fidelity simulation is carried out using high-performance computing in clusters at several parametric points to develop the solution manifolds. The second stage is online usually performed on a less powerful machine such as personal laptops. At this online phase when the parametric space and transient dynamics are learnt using surrogate models such as a data-driven approach, the class of those reduced order approaches is usually termed as a non-intrusive reduced order model. A variety of data-driven ROM techniques have been proposed in recent years, including interpolation-based and machine-learning-based approaches \cite{hajisharifi2025combining, hajisharifi2024comparison, hajisharifi2024lstm}.In heat-transfer applications, non-intrusive ROMs have also been coupled with Bayesian inversion and data assimilation to infer unmeasured boundary fluxes and temperature in real time \cite{bakhshaei2024optimized}.

Gappy data reconstruction implies the completion of the full field variables with partial experimental data at the sensors' location. Several interpolations and least-square estimations have been instrumental in the literature for the approximation of the missing data. Optimal interpolation approach has been used by Reynolds et al. \cite{reynolds1994improved}, Smith et al. \cite{smith1996reconstruction}, Kaplan et al.\cite{kaplan1998analyses}. The kriging approach \cite{oliver1990kriging} is used for the evaluation of the missing points using local weighted averaging. Everson and Sirovich \cite{everson1995karhunen} proposed an approach combining the Proper Orthogonal Decomposition (POD) with the least-square estimation for the reconstruction of "gappy" or missing data. The gappy POD approach has been used for several applications for the reconstruction of the full flowfield using experimental and numerical data at the sensor points such as in the case of the aerospace application by Bui-Thanh et al. and Wilcox \cite{bui2004aerodynamic, willcox2006unsteady}, cavity flow \cite{murray2007application}, boundary layers \cite{gunes2006gappy}, biomedical application \cite{yakhot2007reconstruction}. An algorithm based on frequency-domain, or spectral proper orthogonal decomposition (SPOD), for gappy data reconstruction, was proposed by Nekkanti et al. \cite{nekkanti2023gappy}. Recently, several deep-learning-based approaches have been considered for the reconstruction of the flow field using nonlinear methods like convolution neural networks (CNN) \cite{sekar2019fast} and generative adversarial networks \cite{yousif2023deep}. However, there are limited works on the employment of such a deep-learning approach for the use of both high-fidelity numerical data and experimental data (at the sensor locations).  

In this study, we introduce a novel non-intrusive reduced-order modeling (ROM) approach that integrates data-driven techniques like POD-ANN (Proper Orthogonal Decomposition combined with Artificial Neural Networks) with the traditional Gappy POD method. This approach is termed as ANNGPOD. This hybrid framework reconstructs the full-field variable using numerical simulations at multiple parametric points alongside sparse sensor data. The proposed method exhibits strong robustness and delivers accurate predictions of the temperature field, irrespective of the number or placement of the sensor points. Furthermore, Proper Orthogonal Decomposition coupled with Radial Basis Function (POD-RBF) based ROM approach is also applied to the refrigerator CFD model to explore the parametric space and compared with the proposed ANNGPOD method. Here, only the high-fidelity numerical data set is considered.

We organise the current work in the following way: Section \ref{sec:FOM} describes the full-order CFD model associated with the domestic fridge model. The reduced order model based on the POD-RBF approach, conventional GPOD method and our proposed deep learning enhanced GPOD method termed ANNGPOD are described in Section 
 \ref{sec:ROM}. Finally, the numerical results associated with the full-order model and reduced order model are discussed in Sections \ref{sec:res} and \ref{sec:Rom_res}. At the end of Section \ref{sec:Rom_res} we have introduced the computational speed-up due to different ROM approaches.  

\section{The full order model}\label{sec:FOM}

A conjugate heat transfer (CHT) algorithm is used to carry out the steady-state Reynolds-Averaged Navier-Stokes (RANS) simulations. This method is based on solving the conservation of mass, momentum and energy in fluid and the conservation of energy in solid, simultaneously \cite{zhang2014conjugate, kim2020simulation, laitinen2020computational, godino2022cfd, lahaye2022modeling}. To ensure the continuity of temperature, the fluid and solid region are coupled by exchanging the information at the fluid-solid, solid-solid interfaces. The governing equations in the fluid region are described in Sec. \ref{sec:fluid}, while the governing equation in the solid region is presented in  Sec. \ref{sec:solid}.

%%%%%%%%%%%%%%%%%%%%%%%%%%%%%%%%%%%%%%%%%%%%%%%%%%%%%%%%%%%%%%%%%%%%%%%%%%%%%%%%%%%%%%%%
%%%%%%%%%%%%%%%%%%%%%%%%%%%%%%%%%%%%%%  CFD  %%%%%%%%%%%%%%%%%%%%%%%%%%%%%%%%%%%%%%%%%%%%%%
%%%%%%%%%%%%%%%%%%%%%%%%%%%%%%%%%%%%%%%%%%%%%%%%%%%%%%%%%%%%%%%%%%%%%%%%%%%%%%%%%%%%%%%%

\subsection{Governing equations for fluid}\label{sec:fluid}

The steady-state motion of compressible fluid in a spatial domain of interest $\Omega_f$ is described by Navier-Stokes equations:

\begin{align}
% \centering
&\frac{\partial \rho u_j}{\partial x_j} = 0 &&\text{in} \ \Omega_f, \label{continuity} \\
&\frac{\partial}{\partial x_j}(\rho u_i u_j) = -\frac{\partial p'}{\partial x_i} + \frac{\partial}{\partial x_j}(\mu \frac{\partial u_i}{\partial x_j}) + \frac{\partial}{\partial x_j}(-\rho u'_i u'_j) &&\text{in} \ \Omega_f , \label{momentu} \hspace{1cm} 
\end{align}

\noindent where $\rho$ is density, $u$ represent the mean velocity and $u'$ represent the fluctuating velocity. $\mu$ is the dynamic viscosity and $p'$ is the modified pressure, defined as:

\begin{equation}
    p' = p + \frac{2}{3} \rho k + \frac{2}{3}(\mu + \mu_t) \frac{\partial u_k}{\partial x_k},
    \label{modified_pressure}
\end{equation}

where $p$ is the pressure, $\mu_t$ is the eddy viscosity and $k$ turbulent kinetic energy. The subscript $i$, $j$ and $k$ denote the velocity component in the $x_i$, $x_j$ and $x_k$ directions, respectively. The last term in \ref{momentu}, represents Reynolds stress tensor,$\tau_{ij}$, which is defined as follows:

\begin{equation}
\tau_{ij} = \rho u'_i u'_j = -2 \mu_t \left( S_{ij} - \frac{1}{3} \frac{\partial u_k}{\partial x_k} \delta_{ij} \right) + \frac{2}{3} \rho k \delta_{ij},
\end{equation}
where 
\begin{equation}
    S_{ij} = \frac{1}{2} \left( \frac{\partial u_i}{\partial x_j} + \frac{\partial u_j}{\partial x_i} \right)
\end{equation}

In this study, we employed a two-equation eddy-viscosity turbulence model where $\mu_t$ is determined using $k-\omega$ SST model. In this model, the following two transport equations are solved for the dissipation rate, denoted as $\omega$, and turbulent kinetic energy, $k$, \cite{menter1994two, rumsey2010compressibility, wilcox2008formulation, bakhshaei2021multi}:

\begin{equation}
    \frac{\partial (\rho u_j k)}{\partial x_j} = \mathcal{P} - \beta ^* \rho \omega k + \frac{\partial}{\partial x_j} \left [  (\mu + \sigma_k \mu_t) \frac{\partial k}{\partial x_j}  \right],
    \label{eq:K_eq}
\end{equation}

\begin{equation}
    \frac{\partial (\rho u_j \omega)}{\partial x_j} = \frac{\gamma \rho}{\mu_t}   \mathcal{P} - \beta \rho \omega^2  + \frac{\partial}{\partial x_j} \left [  (\mu + \sigma_{\omega} \mu_t) \frac{\partial \omega}{\partial x_j}  \right],
    \label{eq:omega_eq}
\end{equation}

where the constants are $\beta^* = 0.09$, $\sigma_k=0.5$, $\gamma=5/9$, $\beta=3/40$ and  $\sigma_{\omega}=0.5$.  The production term $\mathcal{P}$  is given by :
\begin{equation}
     \mathcal{P} = -\tau_{ij} \frac{\partial u_i}{\partial x_j}
    \label{eq:omega_eq}
\end{equation}

Given the significant temperature difference between the upper and lower parts of the computational domain considered in this study, we adopted a compressible ideal gas model. The energy equation for fluid region reads as:

\begin{align}
    &\frac{\partial}{\partial x_i} \left(u_i (\rho \bm{E} + p) \right) = \frac{\partial}{\partial x_j} \left( \alpha_f \frac{\partial T_f}{\partial x_j} + u_i \frac{\partial}{\partial x_j} \left (  \mu \frac{\partial u_i}{\partial x_j} - \rho u'_i u'_j  \right ) \right) + Q_{\text{r}} &&\text{in} \ \Omega , 
    \label{eq:Energy}
\end{align}

where $\bm{E}$ is the total energy, $\alpha_f$ the fluid thermal diffusivity, $T_f$ temperature in the fluid region and $Q_{\text{r}}$ the heat source or sink due to the presence of radiation. Radiation, which has been demonstrated to play an important role in CHT simulations \cite{zhang2014conjugate, balaji1994correlations, laguerre2007numerical, bayer2013cfd}, was modeled in our analysis by using Surface-to-Surface (S2S) model \cite{frank2016modeling,ben2017cfd}. This model approximates the expensive computation of radiation between surfaces to cheaper calculations. S2S model partitions the participating surfaces into discrete sections known as \textit{view factors}. It computes the radiative heat flux by considering the temperature and emissivity of the surfaces,  as well as their geometrical relationship. In this model, each surface is treated as both a source and sink of radiation, with its emissivity and absorptivity assumed to be equal and constant \cite{frank2016modeling}. The net energy leaving surface $A$ due to radiation can be defined as the sum of radiative heat flux of each discrete surface, $A_k$, as follows:

\begin{equation}
    Q_{\text{r}} = \sum_k \int_{A_k} q_k \, dA, 
\end{equation}

where $A_k$ is the area of surface $k$. The radiative heat flux leaving surface $k$, denoted as $q_k$, is computed as follows \cite{incropera1996fundamentals, mazumder2019application,burns1998surface} :

\begin{equation}
    q_k= \epsilon_k \sigma T_{s,k}^4 + \mathcal{R}_k  \displaystyle\sum_{j=1}^N F_{k,j} q_j,
\end{equation}

% \begin{equation}
%     \frac{\sigma(T_i^4 - T_j^4)}  { \frac{1-\epsilon_i}{\epsilon_i A_i} + \frac{1}{A_i F_{ij}} + \frac{1-\epsilon_j}{\epsilon_j A_j} }
% \end{equation}

where $q_j$ represents the radiative heat flux
from the surroundings surface, $\sigma =  5.67 \times 10^{-8} \; w/(m^2K^4)$ is the Stefan-Boltzmann constant, $T_{s,k}$ is the temperature of surface $k$, $\epsilon_k$ and $\mathcal{R}_k$ represent the surface emissivity and reflectivity, respectively. The view factor $F_{k,j}$ indicates the fraction of radiation that leaves surface $k$ and hits surface $j$. This factor is a function of the geometric configuration and orientation of the surfaces \cite{buschgens2022radiation,ben2017cfd}.

\subsection{Governing equations for solid}\label{sec:solid} 

The governing equation for conductive heat transfer in the solid region in a  domain of interest $\Omega$ reads as:
 
\begin{align}
     \frac{\partial}{\partial x_j} \left (\alpha_s \frac{\partial T_s}{\partial x_j}      \right ) = 0 \hspace{3cm}  &&\text{in} \ \Omega , 
    \label{eq:Heat_solid}
\end{align}

where $T_s$ is the temperature in solid region  and $\alpha_s = \lambda/(\rho C_p)$ is the thermal diffusion coefficient. $\lambda$, $\rho_s$  and $C_p$ are thermal conductivity, density, and specific heat of solid, respectively. 

We used the second-order accurate finite volume method (FVM), implemented in  the open-source C++ library OpenFOAM\textsuperscript{\textregistered}\cite{weller1998tensorial}, to discretize the partial differential equations. The Semi-Implicit Method for Pressure Linked Equations (SIMPLE) algorithm was used to couple pressure and velocity  \cite{patankar1983calculation}. In this algorithm, a prediction step followed by a correction step is used to couple pressure and velocity fields.

\section{\bf{Reduced order model}} \label{sec:ROM}

%GS comments here : 

% GS Note: this sentence is not fully clear, what do you mean?
% GS Note: this is also true for an intrusive ROM
% GS Note: we should first introduce the acronym
% GS Note: ANN is not an interpolation approach, better to use regression
% GS Note: introduce before the acronym
% GS Note: I see that some acronyms are introduced in the abstract. Better to introduce them also the first time they appear in the paper

In this section, we introduce a novel non-intrusive parametric reduced-order model that reconstructs the temperature fields at regions of interest in a domestic refrigerator, leveraging high-fidelity numerical data at known parameter values together with sparse experimental measurements from sensors at an unknown parameter value. A suitable training dataset is the first requirement for developing the reduced-order model. Section~\ref{sec:method_PODI} introduces Proper Orthogonal Decomposition with Radial Basis Function interpolation (POD-RBF), and Section~\ref{sec:ANN} presents its extension with Artificial Neural Networks (POD-ANN). The conventional Gappy Proper Orthogonal Decomposition (GPOD) is then introduced in Section~\ref{sec:method_GPOD}. Next, we introduce ANNGPOD, a stable ANN-augmented GPOD method that extends POD-ANN by enriching its loss function with an additional term derived from the GPOD formulation, as described in Section~\ref{sec:method_ANNGPOD}.

\subsection{POD with Radial Basis Function} \label{sec:method_PODI}

%GS comments here : 
% GS Note: this sentence is not fully clear. Interpolation of subspace?
% GS Note: just a general note. A parameter is different from a parameter value.
% Better to say for different parameter values
% GS Note: I would also introduce the concept of the parameter vector and include the parameter vector dependency into the snapshots matrix GS Note: μi is generally used to denote the i − th parameter
% \GS{what space is $\mathbb{R}^n_{\mu}$? It should go from the dimension of the parameter space to the dimension of the POD space}

We assume that any physical variable, such as temperature in the present case, can be approximated as a linear combination of basis functions $\phi$ (depending only on the spatial variable $\boldsymbol{x}$), weighted by scalar coefficients $\boldsymbol{\alpha}$ that depend on parameters (physical or geometrical) and/or time. Accordingly, the temperature field ($T$) and its reduced approximation ($T_r$) can be expressed as:  

\begin{equation}
\label{eq:linear}
T \approx T_r = \sum_{i=1}^{N_T^r} \boldsymbol{\alpha}_i(t, \boldsymbol{\mu}) \, \phi_i(\boldsymbol{x}),
\end{equation}

where ${N_T^r}$ denotes the cardinality of the reduced basis of the temperature field, while $t$ and $\boldsymbol{\mu}$ represent time and the parameter vector, respectively. In this work, we focus exclusively on parametric variations, where the parameter vector consists of the evaporator temperature $T_{ev}$, ambient temperature $T_{amb}$, and fan velocity percentage $v_f$. Now, to construct the reduced-order model, we employ Proper Orthogonal Decomposition (POD) combined with Radial Basis Function interpolation, known as the POD-RBF approach. The method involves two stages: offline and online. In the offline stage, a linear subspace is obtained via POD along with the corresponding reduced coefficients. Interpolation methods such as those in \cite{demo1,demo2019complete,pier,lvad,hajisharifi2023non} are then used to learn the dependence of the reduced coefficients on the parameters, enabling efficient exploration of the parametric space during the online stage. Complementary to GPOD, Ensemble Kalman–type data assimilation can jointly estimate states and parameters from sparse sensors in fluid systems, improving near-wall or boundary reconstructions \cite{bakhshaei2024stochastic}.

The algorithm begins with a collection of temperature field vectors, referred to as the snapshot matrix $\mathbf{T}$, associated with parameter values $\boldsymbol{\mu}_i$ (see Eqn.~\ref{eq:PODI1}). Here, $N_{dof}$ denotes the number of degrees of freedom of the full-order model, and $M$ is the total number of parameter values considered:  

\begin{equation} \label{eq:PODI1}
\mathbf{T}=
\left[\begin{array}{cccc}
\mid & \mid &        & \mid \\
T_1(\mathbf{x}) & T_2(\mathbf{x}) & \ldots & T_M(\mathbf{x}) \\
\mid & \mid &        & \mid
\end{array}\right] 
\in \mathbb{R}^{N_{dof} \times M}, 
\quad T_i(\mathbf{x}) = T(\mathbf{x}, \boldsymbol{\mu}_i), \quad i=1,\ldots,M.
\end{equation}

where $\boldsymbol{\mu} \in \Omega_{\boldsymbol{\mu}} \subset \mathbb{R}^M$ is the vector containing the associated parameter values $\boldsymbol{\mu}_i$. Applying Singular Value Decomposition (SVD) to the matrix $\mathbf{T}$ yields:  

\begin{equation}
\mathbf{T} = \boldsymbol{U} \boldsymbol{\Sigma} \boldsymbol{V}^T.
\end{equation}

where, $\boldsymbol{U} \in \mathbb{R}^{N_{dof} \times N_{dof}}$, and $\boldsymbol{V} \in R^{M \times M}$ are the matrices comprising of the left and right singular value vectors respectively. $\boldsymbol{\Sigma} \in \mathbb{R}^{N_{dof} \times M}$ is a matrix containing the singular values $\sigma_{i}$. $\boldsymbol{U}_{N_{T}^r}$ is the matrix $\boldsymbol{U}$ truncated to the first $N_{T}^r$ columns representing our POD space. The POD basis $\boldsymbol{U}_{N_{T}^r}$
minimizes $\|\mathbf{T}-\boldsymbol{U}_{N_{{T}^r}} \boldsymbol{U}_{N_{T}^r}^T \mathbf{T}\|_F$ where $\|\|_F$ is the Frobenius norm. 
%\GS{there are some problems here: $\sigma_i$ are singular values and not eigenvalues}
Typically, the value of $N_{T}^r$ is commonly chosen to meet a user-provided threshold $\delta$ for the cumulative energy of the singular values defined as:

\begin{equation} \label{eq:energy_modes}   
\frac{\sum_{i=1}^{N_{T}^r} \sigma_i^2}{\sum_{i=1}^{N_s} \sigma_i^2} \geq \delta.
\end{equation}

After constructing the POD space, we can approximate the input snapshots by using Eqn. \ref{eq:linear}.

\begin{equation}
    T_r\left(\boldsymbol{\mu}_j\right) \approx \sum_{L=1}^{N_{T}^r} \mathbb{A}_L\left( \boldsymbol{\mu}_j\right) \phi_L, \quad \text { with } \quad j=1, \ldots, M,
\end{equation}

where the modal coefficients, $\mathbb{A}_L\left(\boldsymbol{\mu}_j\right)$ are the elements of the matrix $\boldsymbol{C}=\boldsymbol{U}_{N_{T}^r}^T \mathbf{T} \in \mathbb{R}^{N_{T}^r \times N_{dof}}$. Then the ROM is built using as input-output data the pairs $\left\{\left(\boldsymbol{\mu}_j\right), \mathbb{A}_L\left(\boldsymbol{\mu}_j\right)\right\}$ associated to the $L^{th}$ POD mode $\phi_L$ with $L=1, \ldots, N_{T}^r$.

where,

\begin{equation}\label{eq:RBF-POD}
\mathbb{A}_L\left( \boldsymbol{\mu}_j\right)= \sum_{n=1}^{M} w_{L,n} \zeta_{L,n}\left(\left\|\left( \boldsymbol{\mu}_j\right)-\left(\boldsymbol{\mu}_n\right)\right\|\right),
\end{equation}

Here, $\zeta_{L,n}$ are the Radial Basis Functions \cite{buhmann2000radial}, which we chose as Gaussian functions centred in $\left(\boldsymbol{\mu}_n\right)$ and $w_{L, n} \in \mathbb{R}^M$ are unknown weights. Eqn. (\ref{eq:RBF-POD}) can be reformulated in terms of a linear system : 

\begin{equation}
\boldsymbol{Z}_L \boldsymbol{w}_L=\mathbb{A}_L,
\end{equation}

to be solved to obtain the weights $\boldsymbol{w}_L$ for every value of $L=1, \ldots, N_{T}^r$ once at all in the offline phase. Matrix, $\boldsymbol{Z}_L$ consists of the radial basis functions values. 

Then, in the online phase, for any new parameter, $\left(\boldsymbol{\mu}_{\star}\right)$, the approximated modal coefficients, $\mathbb{A}_L\left(\boldsymbol{\mu}^{\star}\right)$, are calculated from Eqn. \ref{eq:RBF-POD} and the ROM solution is computed as: 

\begin{equation}
T_r\left(\boldsymbol{\mu}_*\right)=\sum_{L=1}^{N_{T}^r} \mathbb{A}_L\left(\boldsymbol{\mu}_{\star}\right) \phi_L.
\end{equation}

\subsection{Artificial Neural Network} \label{sec:ANN}

An ANN is a deep learning model consisting of neurons and a set of directed weighted synaptic connections among the neurons. ANN is an oriented graph, with the neurons as nodes and the synapses as oriented edges. The weights associated with the ANN are adjusted by employing a training process to configure the network for a specific application. Let us consider the neuron $j$. Three functions characterize completely the neuron $j$:

\begin{itemize}

\item{the propagation function $u_j$ is used to transport values through the neurons of the ANN. $m$ is the total number of sending neurons linked with the neuron $j$. The weighted sum approach is used where $b_j$ is the bias, $y_{sk}$ is the input related to the sending neuron k, and $w_{sk}$ are the weights.}
\begin{equation}
u_j=\sum_{k=1}^m w_{s_k, j} y_{s_k}+b_j.
\end{equation}

\item{the activation function $f_\text{act}$ which operates on $u_j$ during the training process:

\begin{equation}
a_j=f_{\text {act, j}}\left(\sum_{k=1}^m w_{s_k, j} y_{s_k}+b_j\right).
\end{equation}
Commonly the activation functions are non-linear. Possible choices are sigmoid function, hyperbolic tangent, RELU, SoftMax. More details can be found in \cite{sharma2017activation}.
}

\item{the output function $y_j$ . It is related to the activation function $a_j$ . Often it is the identity function, so that $a_j$ and
$y_j$ coincide:

\begin{equation}
y_j=f_{\text {out }}\left(a_j\right)=a_j.
\end{equation}
}
\end{itemize}
In this work, we will use a specific type of ANN, the feedforward neural network \cite{fine1999feedforward} with multiple layers. In a feedforward neural network, each node in one layer is connected to all nodes in subsequent layers. We highlight that the input layer of our network consists of the set of time/parameter instances $\left\{\left( \boldsymbol{\mu}_1\right), \ldots,\left( \boldsymbol{\mu}_{M}\right)\right\}$ whilst the output one is given by the corresponding modal
coefficients ${\mathbb{A}_L\left(\boldsymbol{\mu}_1\right), \ldots, \mathbb{A}_L\left(\boldsymbol{\mu}_M\right)}$, with $L=1, \ldots, N_{T}^r$. During the training procedure, the weights of the connections in the network are repeatedly changed to minimize the difference between the output vector of the net $\Tilde{\mathbb{A}}_L$ and the required output vector $\mathbb{A}_L$ using backpropagation algorithm \cite{rojas1996backpropagation,rumelhart1986learning}. The key to backpropagation is a method for computing the gradient of the error concerning the weights for a given input by propagating the error backwards through the network. A loss function is introduced to optimize the parameter values in a neural network model. The loss function $\mathcal{L}=\mathcal{L}(\Tilde{\mathbb{A}}_L, \mathbb{A}_L)$ used in this work is the mean squared error (MSE), which is the most common choice for regression problems:

\begin{equation}\label{eq:LPOD_ANN}
\mathcal{L}_{\text{ANN}}=\frac{1}{M}\sum_{i=1}^{M} \mathcal{L}_i=\frac{1}{M}\sum_{i=1}^{M} \frac{1}{L} \sum_{j=1}^L\left(\mathbb{A}_{i, j}-\tilde{\mathbb{A}}_{i, j}\right)^2
\end{equation}

\subsection{Gappy POD} \label{sec:method_GPOD}

The conventional Gappy POD approach computes the temperature field based on temperature sensor data available at a few sensor locations, termed as missing or "gappy" data. A brief description of the Gappy POD approach is described in the current subsection. First, a mask matrix is developed to locate the rows where the data is available. For the temperature field prediction, the corresponding mask vector $l^{k}$ is as follows:

\begin{equation}
\label{eq:GPOD1}
\begin{aligned}
& l^k = 0 \ \ \text{if} \ \ {T^i}(\boldsymbol{\mu}_k) \ \ \text{is missing},\\
& l^k = 1 \ \ \text{if} \ \ {T^i}(\boldsymbol{\mu}_k) \ \ \text{is known},\\
\end{aligned}
\end{equation}

where, ${T^i}(\boldsymbol{\mu}_k)$ denotes the $i^{th}$ element of the vector $T(\boldsymbol{\mu}_k)$, pointwise multiplication is defined as $(l^k,T(\boldsymbol{\mu}_k))^{i}= {l^{i,k}}{T^i}(\boldsymbol{\mu}_k)$. The inner product and induced norm can be expressed respectively as shown in Eqn. \ref{eq:GPOD2}: 

\begin{equation}
\label{eq:GPOD2}
\begin{aligned}
& (z,v)_{l} = ((l,z),(l,v)),\\
& (\Vert v \Vert)^2 = (v,v)_{l}
\end{aligned}
\end{equation}

As discussed in the previous section, the standard POD basis vector is defined as ${\phi}_L$ for the known snapshots set ${T(\boldsymbol{x},\boldsymbol{\mu}_k)}$ where $k = 1 \ \ \text{to} \ \ M$ and $L = 1 \ \ \text{to} \ \ N_{T}^r$.

Now assume, at an unknown parameter, $\mu^*$, partial data is available 
\begin{equation}
\label{eq:GPOD3}
\begin{aligned}
\hat{T}_{\boldsymbol{\mu}_*}=\left[\hat{T}\left(\hat{\boldsymbol{x}}, \boldsymbol{\mu}_*\right)\right] = (l,\left[{T}\left(\boldsymbol{x}, \boldsymbol{\mu}_*\right)\right]) \in \mathbb{R}^{N_{\text {dof.partial }}},
\end{aligned}
\end{equation}

where $\hat{T}$ is the partial temperature data available at locations $\boldsymbol{\hat{x}}$ and parameter $\boldsymbol{\mu}_*$. ${N_{\text {dof.partial }}}$ is the dimension of the sparse dataset available. In the next step, we need to obtain the reduced truncated basis $\Phi_R$ to solve a set of linear equations to minimise the following $\mathcal{L}_{GPOD}$ shown in Eqn. \ref{eq:GPOD4}.

\begin{equation}
\label{eq:GPOD4}
\begin{aligned}
\mathcal{L}_{\text{GPOD}} = \hat{T}_{\boldsymbol{\mu}_*}-\sum_{L=1}^{N_{T}^r} \mathbb{A}_L\left(\boldsymbol{\mu}_{\star}\right) \phi_{R, L}(\hat{\mathbf{x}})
\end{aligned}
\end{equation}

where, $\Phi_R = [{\phi_R}_1, {\phi_R}_2, ...., {\phi_R}_L]$,   ${\phi_R}_i = (l,{\phi}_i)$. $\Phi_R$ is not orthonormal anymore after the selected rows are omitted. If the number of points where data is available and the number of modes are not the same, in that case, the matrix $\Phi_R$ is rectangular. Therefore, Eqn. \ref{eq:GPOD4} is solved by computing the pseudo-inverse of the matrix $\Phi_R$ as shown in Eqn. \ref{eq:GPOD5}.   

\begin{equation}
\label{eq:GPOD5}
\begin{aligned}
\mathbb{A}\left(\boldsymbol{\mu}_{\star}\right)=\left[\Phi_R\right]^{\dagger} \hat{T}_{\boldsymbol{\mu}_*} , \dagger \ \ \text{is the pseudo-inverse of a rectangular matrix}
\end{aligned}
\end{equation}

\subsection{Deep Learning enhanced Gappy POD} \label{sec:method_ANNGPOD}

We propose a deep learning enhanced Gappy POD approach introducing additional loss terms arising from Gappy POD i.e., Eqn. \ref{eq:GPOD4} in the loss term associated with the POD-ANN approach, Eqn \ref{eq:LPOD_ANN}. Now we will describe the inputs and outputs of our proposed method ANNGPOD. The Input layer of ANNGPOD consists of the set of parameter values $\left\{\left( \boldsymbol{\mu}_1\right), \ldots,\left( \boldsymbol{\mu}_{M}\right)\right\}$. The objective is to reconstruct the temperature field at $\boldsymbol{\mu}_{*}$ where sparse measurements of temperature $\hat{T}_{*}$ are available which acts as additional input to the ANNGPOD network. The output is given by the corresponding modal
coefficients ${\mathbb{A}_L\left(\boldsymbol{\mu}_1\right), \ldots, \mathbb{A}_L\left(\boldsymbol{\mu}_M\right)}$, with $L=1, \ldots, N_{T}^r$. Following are the steps of the ANNGPOD. 

\begin{itemize}
\item{At every epoch iteration, the prediction of the POD-ANN network at $\boldsymbol{\mu}_{*}$ is $\tilde{\mathbb{A}}_{i,j=*}$ with $i=1, \ldots, N_{T}^r$}
\item{The temperature field can be computed from the $\tilde{\mathbb{A}}_{i,j=*}$ as follows:}
\begin{equation}    
\tilde{T}_r\left(\boldsymbol{\mu}_*\right) = \sum_{i=1}^{N_{T}^r} \tilde{\mathbb{A}}_i\left(\boldsymbol{\mu}_{\star}\right) \phi_i
\end{equation}

\item{Additional loss term arising from sparse measurement data at location $(\hat{\boldsymbol{x}}_i)$ can be computed as:}

\begin{equation}
\mathrm{L}_{\text {GPOD}} = \frac{1}{N_{\hat{\boldsymbol{x}}}}\sum_{i=1}^{N_{\hat{\boldsymbol{x}}}}(\hat{T}_{\boldsymbol{\mu}_*}(\hat{\boldsymbol{x}}_i)-\tilde{T}_r\left(\boldsymbol{\mu}_*, \hat{\boldsymbol{x}}_i\right))^2
\end{equation}

where, ${N_{\hat{\boldsymbol{x}}}}$ is the total number of sensors where measurement data is available.

\item{The total loss term associated with the ANNGPOD network which goes into the backpropagation algorithm to update the weight matrix and bias vectors in the next epoch iteration is as follows:}

\begin{equation}\label{eq:anngpod}
\mathrm{L}_{\text {ANNGPOD}} =\lambda_1\frac{1}{M}\sum_{i=1}^{M} \frac{1}{L} \sum_{j=1}^L\left(\mathbb{A}_{i, j}-\tilde{\mathbb{A}}_{i, j}\right)^2 + \lambda_2\frac{1}{N_{\hat{\boldsymbol{x}}}}\sum_{i=1}^{N_{\hat{\boldsymbol{x}}}}(\hat{T}_{\boldsymbol{\mu}_*}(\hat{\boldsymbol{x}}_i)-\tilde{T}_r\left(\boldsymbol{\mu}_*, \hat{\boldsymbol{x}}_i\right))^2
\end{equation}
The constants, $\lambda_1$ and $\lambda_2$ are associated with the loss term arising from the POD-ANN and GPOD method, respectively.
\end{itemize}

We explain the steps of the ANNGPOD method in the following pseudo-code. Furthermore, We like to mention that POD-RBF approach mentioned in Section \ref{sec:method_PODI} can be coupled with the Gappy POD to improve the performance of the conventional Gappy POD method, however, in ANN or any deep learning framework, loss terms can be easily modified and such flexibility in the implementation step has motivated the choice of the current regression method, POD-ANN to demonstrate the proposed approach for a practical industrial applications. 

\SetKwComment{Comment}{/* }{ */}

\begin{algorithm}[hbt!] 
\caption{pseudo-code for ANNGPOD Algorithm}\label{alg:algo}
\KwData{$\boldsymbol{u}=\left\{\left( \boldsymbol{\mu}_1\right), \ldots,\left( \boldsymbol{\mu}_{M}\right)\right\}$, ${\mathbb{A}_L\left(\boldsymbol{\mu}_1\right), \ldots, \mathbb{A}_L\left(\boldsymbol{\mu}_M\right)}$, with $L=1, \ldots, N_{T}^r$, $\tilde{T}_{\boldsymbol{\mu}_*}$}

$[{W},{b}] \gets \textbf{INIT} ([{W},{b}])$ \Comment*[r]{Initialize Weight and bias}
\While{$\text{epoch} \leq \text{Total Epoch No.}$}{
  
  $\tilde{\mathbb{A}}_{i,j} , i =1, L, j =1,M \gets \text{NN}({W},{b},f_\text{act}, \textbf{u})$ \Comment*[r]{prediction from chosen network $\text{NN}$}
  $\mathrm{L}_{\text {POD-ANN }} \gets \frac{1}{M}\sum_{i=1}^{M} \frac{1}{L} \sum_{j=1}^L\left(\mathbb{A}_{i, j}-\tilde{\mathbb{A}}_{i, j}\right)^2$ \Comment*[r]{compute POD-ANN based loss term}
  
  $\tilde{\mathbb{A}}_{i,j=*} \gets \text{NN}({W},{b},f_\text{act},\boldsymbol{\mu}_{*})$ \Comment*[r]{prediction from the net at parameter $\boldsymbol{\mu}_*$}

  $\tilde{T}_r\left(\boldsymbol{\mu}_*\right) \gets \sum_{L=1}^{N_{T}^r} \tilde{\mathbb{A}}_L\left(\boldsymbol{\mu}_{\star}\right) \phi_L
  $ \Comment*[r]{Reconstruct the temperature field at $\boldsymbol{\mu}_*$}
  
  $\mathrm{L}_{\text {GPOD}} \gets \frac{1}{N_{\hat{\boldsymbol{x}}}}\sum_{i=1}^{N_{\hat{\boldsymbol{x}}}}(\hat{T}_{\boldsymbol{\mu}_*}(\hat{\boldsymbol{x}}_i)-\tilde{T}_r\left(\boldsymbol{\mu}_*, \hat{\boldsymbol{x}}_i\right))^2$
\Comment*[r]{Compute the loss term based on GPOD}

  $L \gets (\mathrm{L}_{\text {POD-ANN }} + \mathrm{L}_{\text {GPOD}})$ \Comment*[r]{compute total loss term for backpropagation}

  $\Delta {W} \leftarrow-\alpha \mathrm{G}_{\mathrm{ADAM}}\left(\nabla_{{W}} L\right), \Delta {b} \leftarrow-\alpha \mathrm{G}_{\mathrm{ADAM}}\left(\nabla_{{b}} L\right)$ \; \Comment*[r]{compute weight matrices and bias vector update}
  ${W} \leftarrow {W}+\Delta {W}, {b} \leftarrow {b}+\Delta {b}$ \;
}
\end{algorithm}

\newpage 
\section{Numerical results}\label{sec:res}
\graphicspath{{./img_CFD/}}

\begin{figure} %\hspace{20cm}
\centering
  \includegraphics[width=110mm,scale=1.0]{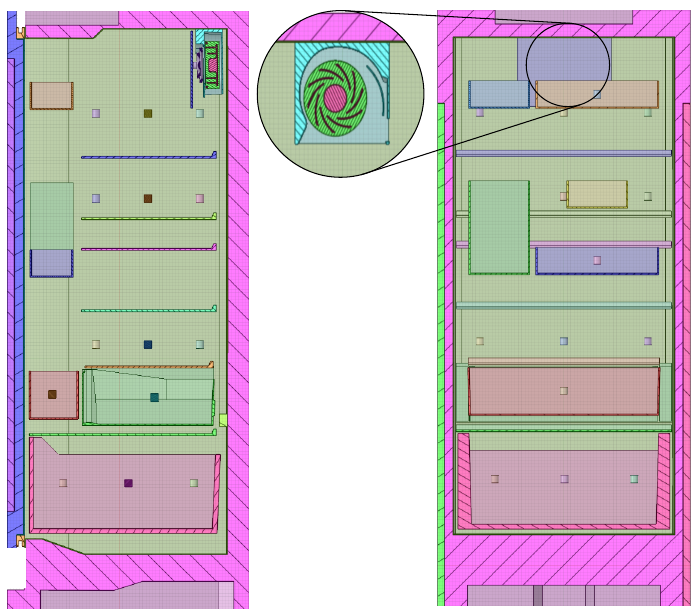}
        \caption{Sketch of the computational domain: the left and right panels illustrate the side and front views of the geometry. The sub-panel highlights the fan region and its configuration.   }
\label{fig:geometry}
\end{figure}

In this work, we studied a household fridge with dimensions $1205 \times 510 \times 485$ mm. 
\fig{fig:geometry} illustrates the side and front views of the fridge under investigation on the right and left panels, respectively. The fridge includes the following components essential for its functionality: door bins which are located on the fridge door, a crisper and a drawer positioned at the bottom of the fridge to store food items with humidity controls for preservation. 
Moreover, there are liners and gaskets designed to contribute to the insulation by creating an airtight closure. The foam insulation surrounds the inner cavity to minimize the heat transfer to and from the environment. Furthermore, there are 4 shelves to separate the inner cavity space and a fan located at the top of the fridge.  The fan is covered by a frame as highlighted in the sub-panel of \fig{fig:geometry}. The thermal behavior of the fridge is monitored by 26 temperature sensors located in various locations of the fridge. The sensors are modeled as solid cylinders with a height of $12$ mm and a radius of $5$ mm.

The entire domain is divided into fluid and solid regions to form the computational domain for CHT simulation. The fluid region represents the air inside the fridge compartment, whereas the solid regions represent the various components of the fridge such as the door bins, crisper, drawer, sensors and others. Different material properties like density, specific heat, and thermal conductivity are defined for each region to accurately model the thermal behavior of each region. Solid-fluid or solid-solid regions interact thermally with each other by exchanging information through interfaces.

The convective heat transfer boundary condition was applied to the external surfaces of the fridge to model the heat transfer between the outer solid and the environment fluid, using a fixed heat transfer coefficient, $h= 4 \ W/(m^2 K)$. The temperature coupling boundary condition was applied to both fluid-solid and solid-solid interfaces. This boundary condition links the temperature of adjacent domains at their interfaces to accurately model the heat transfer in CHT simulations. The no-slip boundary condition was imposed on all the interfaces between air and solid components. Furthermore, we set the fixed temperature boundary condition for the fridge evaporator. This component acts as the cooling source to lower the temperature of its adjacent solid, which is the inner liner of the fridge. Consequently, the inner liner, that surrounds the air, cools down the air. 
%%%%%%%%% Mesh 
An accurate CHT simulation requires a conformal multi-region mesh. The multi-region mesh, where separate meshes are generated for each region, is necessary for CHT  simulations as it provides the flexibility to solve specific types of equations for each region and assign specific boundary conditions to different interfaces.  The conformal mesh guarantees that there are no gaps or overlaps between regions which enables us to have a more accurate calculation of temperature and heat fluxes at the interface.
\begin{figure}[t]
    \centering
        \includegraphics[width=90mm,scale=0.5]{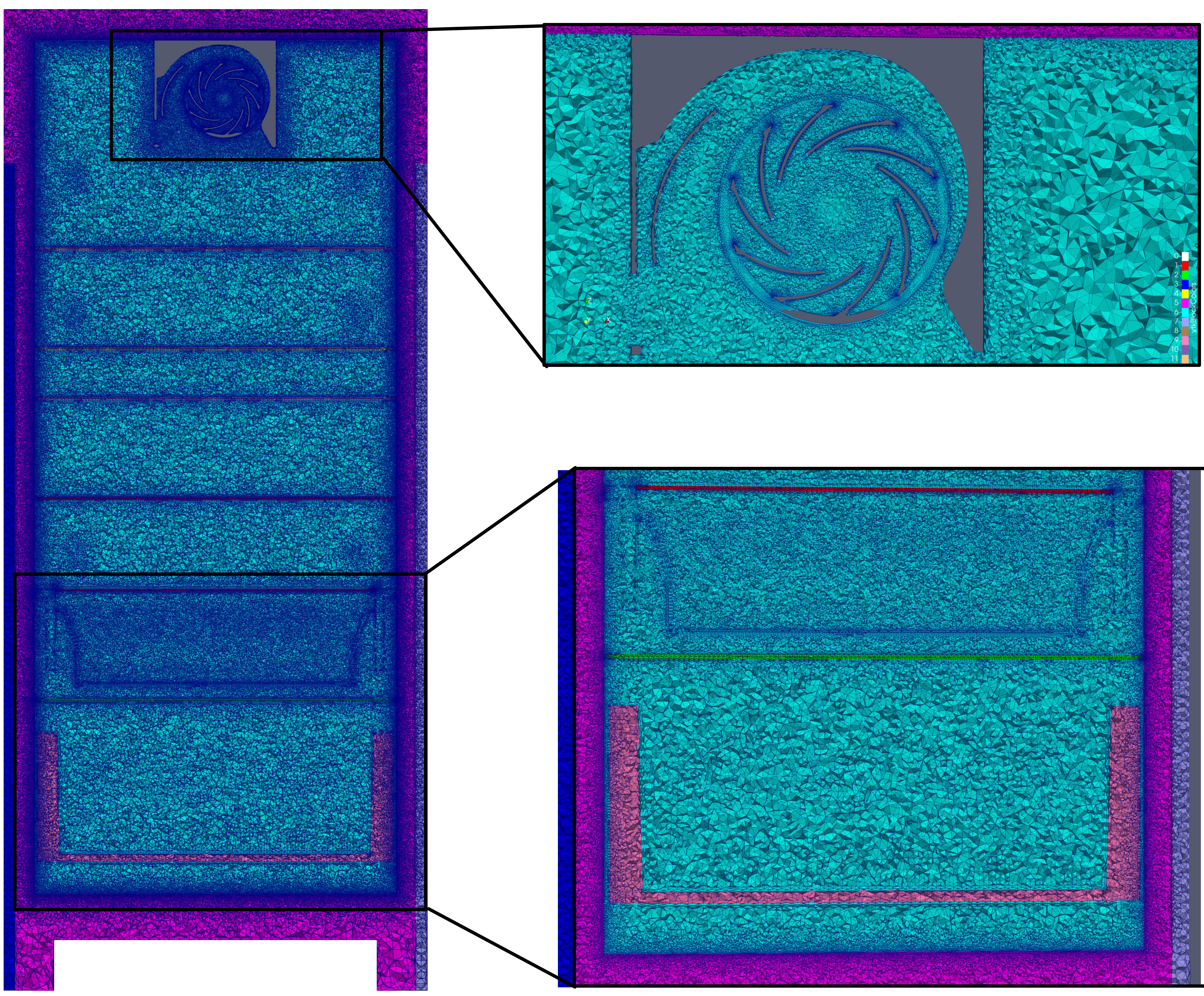}
        \caption{Mesh visualization: the figure displays the mesh of the computational domain. The two sub-panels provide a detailed view of the mesh in the fan region and the bottom part of the fridge.}
        \label{fig:Mesh}
\end{figure}

A sketch of the obtained multi-region mesh is illustrated in \fig{fig:Mesh}. The mesh consists of tetrahedral, hexahedral and polyhedral cells with the minimum and maximum mesh sizes of $0.1$ mm and $32$ mm, respectively, which results in a total of  $20$ millions of elements, approximately, for this geometry. Since the only equation to solve in solid components is the conduction equation, the mesh size is larger in these regions than the fluid region. The zoom-in sub-figures of \fig{fig:Mesh} highlight the local refinements in the region of higher gradients such as near interface regions and fan. The refinement is essential to ensure that high gradients are accurately captured. The max non-orthogonality factor of the obtained mesh is less than 60 $^{\circ}$, demonstrating that the resulting mesh is of good quality for CHT simulations. 

%%%%%% Modeling the fan 
The fan is modeled using the multiple reference frame (\textit{MRF}) method \cite{peng2019strategy}. This approach is a commonly used technique for fan modeling that provides a trade-off between accuracy and efficiency in simulations. In this approach, the fluid region is divided into two different regions: stationary and moving regions. \textit{MRF} models the relative motion between the two regions by including a source term without actually moving the mesh. 

% \begin{table}[h]
% \centering
%  \begin{tabular}{||c |c |c |c||} 
%  \hline
% Max aspect ratio  & Mesh non-orthogonality max &  Mesh non-orthogonality average  & max skewness  \\ [0.5ex] 
%  \hline\hline
% 23.95 & 59.96 & 7.27 & 3.9 \\ %[1ex] 
%  \hline
%  \end{tabular}
%  \caption{Mesh statistics: the four key values of mesh quality in CHT simulations. }
%  \label{tab:Mesh_Statistics}
% \end{table}

\subsection{Validation}
To ensure the accuracy of our model, we validated the results of the CHT simulation against the experimental setup. For this purpose,  two different fridge configurations were considered: static and ventilated fridges. In the static setup, the fan is inactive and the flow is driven by natural convection where buoyancy governs the fluid motion. On the other hand, in the ventilated configuration, the fan works with its maximum angular velocity  $\omega = 191.6$ (rad/sec) resulting in forced convection. For both setups, we set the evaporator temperature ($T_{evap}$) to -$15 \; ^{\circ}C$ and ambient temperatures $(T_{amb})$ to $32 \; ^{\circ} C$. All the solid regions and air were initialized with the ambient temperature $T_{initial}=T_{amb} = 32 \; ^{\circ} C$ to simulate a scenario where the fridge operates during warmer seasons. 
The steady-state temperature at the sensor points were recorded and compared against experimental measurements. To consider the thermal inertia of the solid in our study, we registered the temperature in the middle point of the cylindrical-shaped solid sensors rather than the cylinder surface which results in more accurate measurement.

\begin{figure}[t]
    \centering
        \includegraphics[width=60mm,scale=0.5]{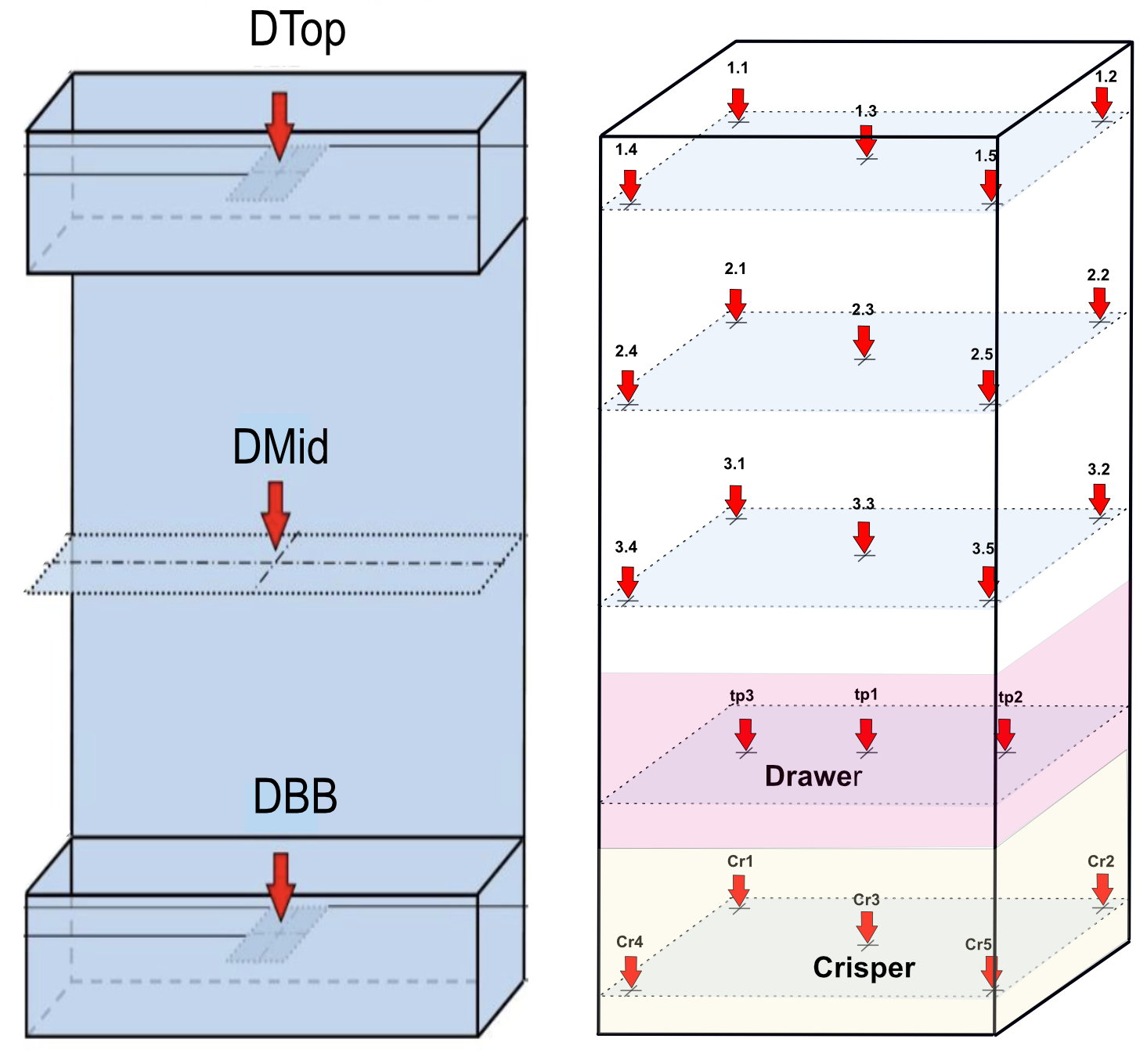}
        \caption{Sensor locations: The right panel illustrates the position of the sensor within the fridge cabin, while the sensor locations in the fridge door are shown in the left panel. }
        \label{fig:sensor_loc}
\end{figure}

\fig{fig:sensor_loc} illustrates the locations of the sensors in the cabin and on the door of the fridge. The sensors located at the upper part of the fridge, with 5 sensors above each shelf, were named using two digits: the first digit indicates the level where the sensor is located, while the second one defines the sensor number at that specific level. There are also 8 sensors in the bottom part of the fridge; 3 sensors in the drawer and 5 in the crisper. As the drawer volume is smaller than the crisper, fewer sensor numbers are used to monitor the temperature in the drawer. The locations of these 23 sensors are shown in the right panel of \fig{fig:sensor_loc}. Moreover, the left panel of \fig{fig:sensor_loc} demonstrates the location of 3 sensors in the top, middle and bottom bins of the fridge door. 

\begin{figure}[t]
    \centering
        \includegraphics[width=160mm,scale=0.5]{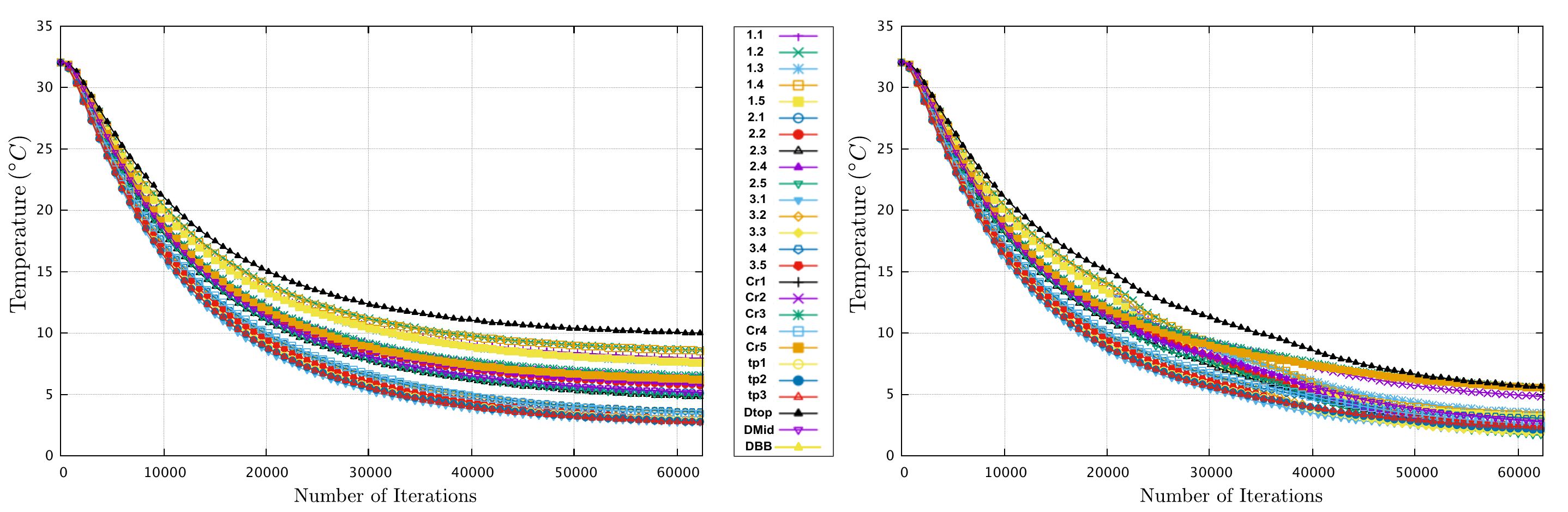}
        \caption{Steady state analysis: time evolution of the temperatures at sensor points. The left and right panels depict the temperature variation over time in the static and ventilated fridge setup, respectively.}
        \label{fig:stationary analysis}
\end{figure}

For both cases, we let the fluid flow evolves until the steady state in temperature was reached  for the all sensors. The time evolution of temperature at the sensor points is depicted in \fig{fig:stationary analysis} with the right and left panels showing the static and ventilated fridge configurations, respectively. 
Since both fridge configurations were initialized with the ambient temperature, all the sensors registered $T = 32 \; ^{\circ} C$ as the initial temperature. 
As the simulations progressed, the temperature at sensor points decreases due to the heat transfer in the fridge until  stabilizing at a plateau. The static setup reached the stationary state after $50,000$ iterations, while the achievement of steady-state was delayed for $5,000$ more iterations in ventilated configuration due to the more complex dynamics of the fluid flow. We continued the simulation until $62,000$ iterations in both cases to ensure the attainment of steady-state conditions. 

For both cases, we observed that the highest temperature was recorded at the "\textit{Dtop}" sensor, located at the top bin of the fridge door. We noticed that this temperature is lower in the ventilated case compared to the static setup, which is aligns with our expectations. The reason is related to the enhanced mixing of the fluid flow caused by the presence of the fan. Consequently, the more uniform temperature distribution in the ventilated setup reduces temperature stratification between different levels of the fridge.

\begin{figure}[ht!]\centering
\subfloat[Static Fridge]{\label{fig:vel_Static}\includegraphics[width=.45\linewidth]{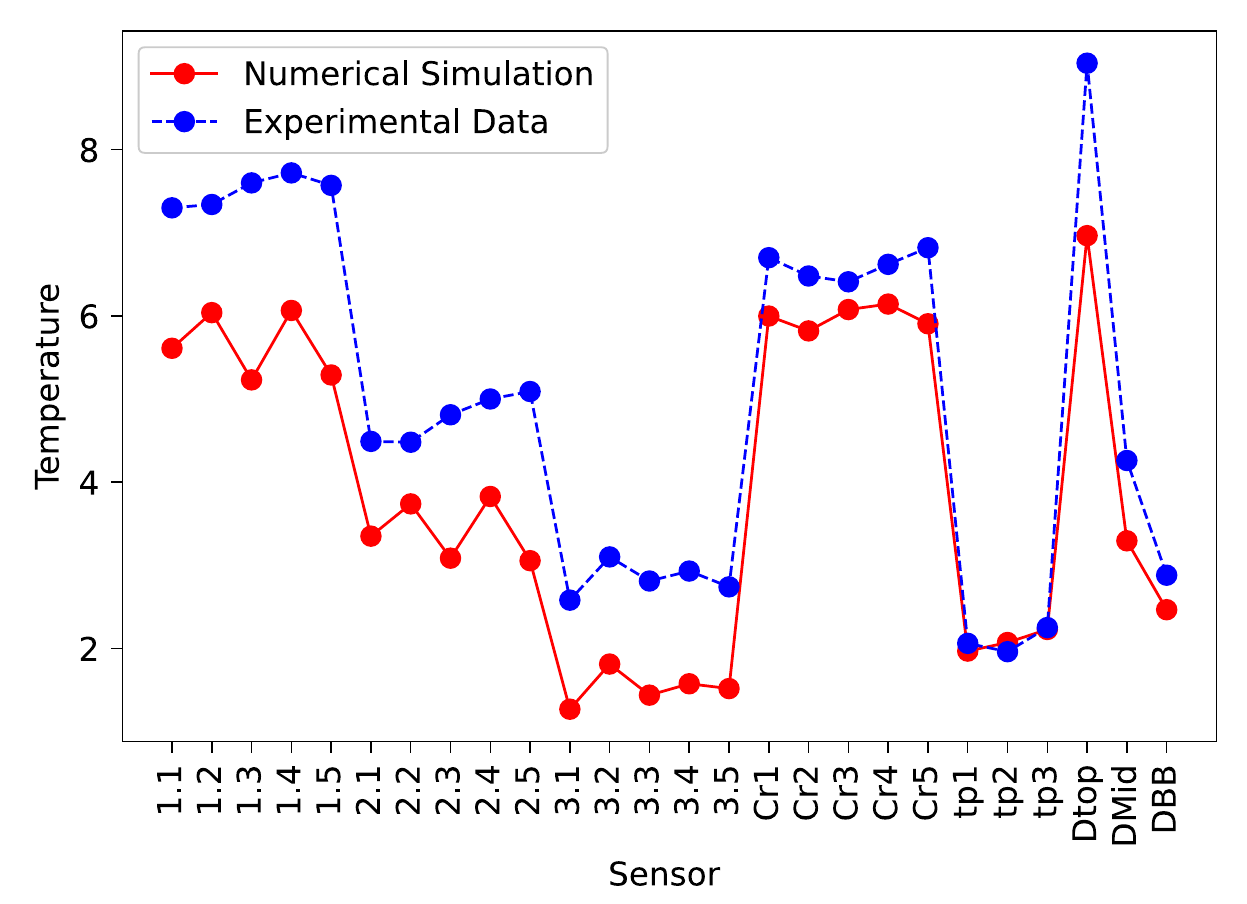}} \hspace{1cm}
\subfloat[ventilated Fridge]{\label{fig:vel_vent}\includegraphics[width=.45\linewidth]{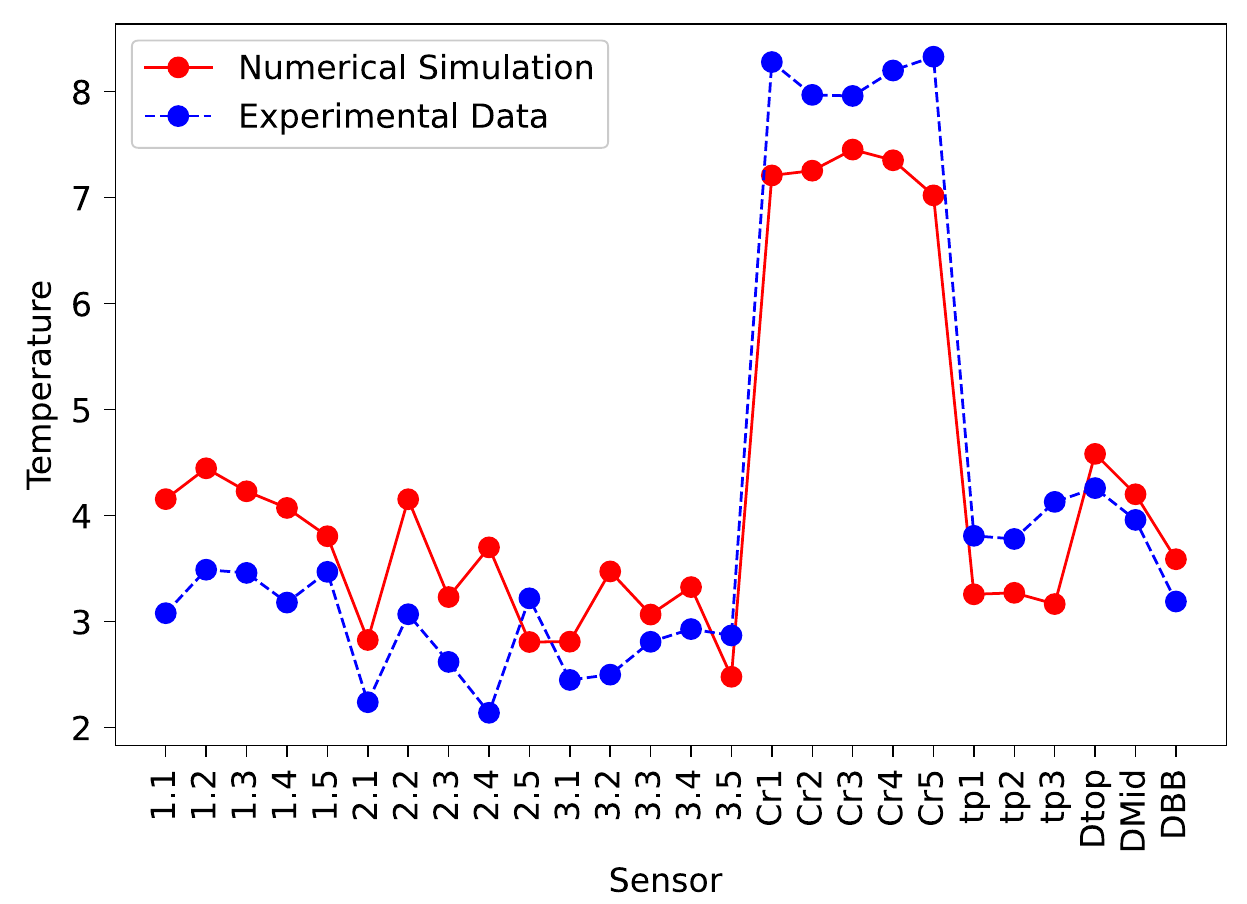}} 
\caption{Validation of CHT model against experimental data. Left and right panels present the validation for static and ventilated fridge configurations, respectively. }
\label{fig:validation}
\end{figure}

The comparisons of the steady-state temperature at the sensor points against experimental measurement were depicted in \fig{fig:validation} with the right and left panels showing the validation in static and ventilated fridge configurations, respectively. In both cases, the numerical data follow the same trend as the experiment. The maximum average temperature difference in both setups was observed at the sensors located on the first level of the fridge, $\Delta T_{avg}=1.7 \; ^{\circ} C$ and  $\Delta T_{avg}=0.78 \; ^{\circ} C$ in static and ventilated configurations, respectively. The average temperature difference in the rest of the fridge is less than $1.5 \; ^{\circ} C$ which is an acceptable value for both cases.

%%%%%%%%%%%%%%%%%%%%%%%%%%%%%%%%%%%%%%%%%%%%%%%%%%%%%%%%%%
%%%%%%%%%%%%%%%%%%%%%%%%%%%%%%%%%%%%%%%%%%%%%%%%%%%%%%%%%%
%%%%%%%%%%%%%%%%%%%%% till here %%%%%%%%%%%%%%%%%%%%%%%%%%
%%%%%%%%%%%%%%%%%%%%%%%%%%%%%%%%%%%%%%%%%%%%%%%%%%%%%%%%%%%%%%%%%%%%%%%%%%%%%%%%%%%%%%%%%%%%%%%%%%%%%%%%%%%%%%%%%%%%

\begin{figure}[t]
    \centering
        \includegraphics[width=75mm,scale=0.7]{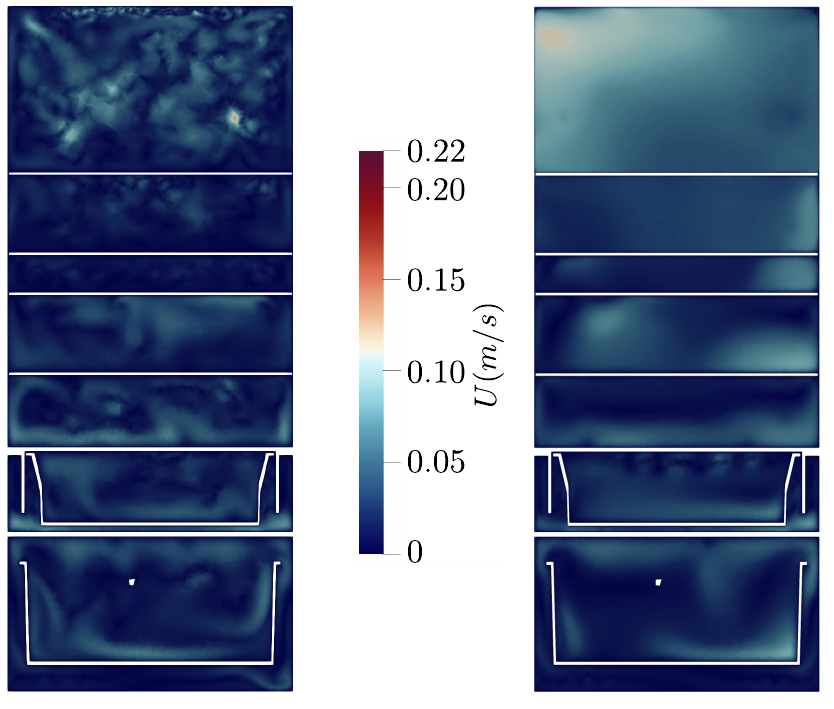}
        \caption{Velocity contour: visual comparison of the magnitude of velocity distribution in the static (left panel) and ventilated fridge, (right panel). }
        \label{fig:vel_contour}
\end{figure}

\begin{figure}[t]
    \centering
        \includegraphics[width=75mm,scale=0.9]{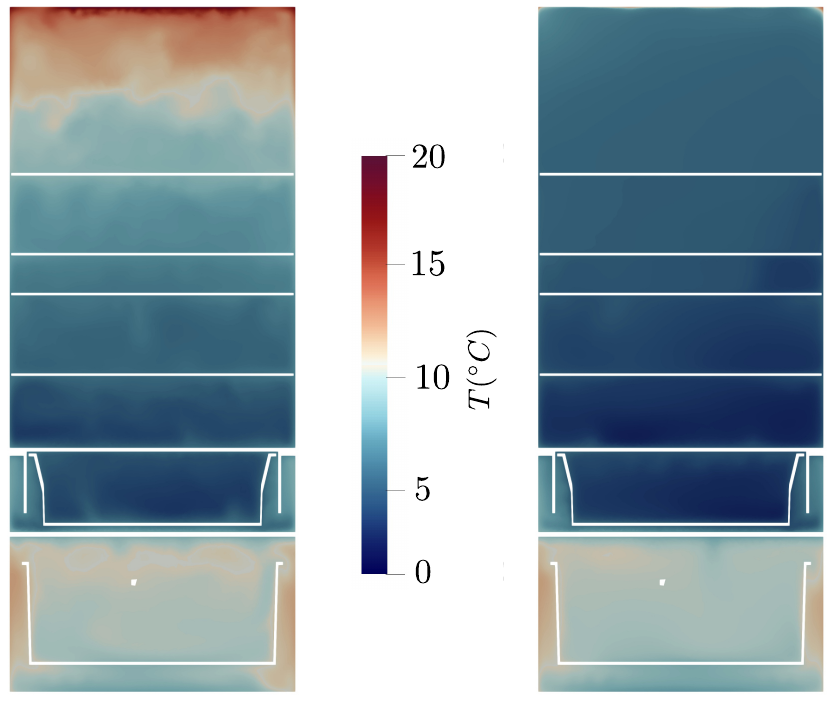}
        \caption{Temperature contour: visual comparison of temperature distribution in the static (left panel) and ventilated fridge, (right panel). }
        \label{fig:temp_contour}
\end{figure}

The visual comparisons of velocity and temperature distribution on the side view for both static and ventilated configurations are presented in \fig{fig:vel_contour} and \fig{fig:temp_contour}. In both figures, the left and right panels display the velocity and temperature distributions in the static and ventilated fridge, respectively.
 \fig{fig:temp_contour}  clearly demonstrates that there is temperature stratification in the static fridge, where the upper region of the fridge is warmer than the lower part. Conversely, the ventilated fridge exhibits an improvement in temperature distribution; the temperature is more uniform and there is no significant temperature difference between the upper and lower parts of the fridge as a result of enhanced mixing.  The velocity contours in \fig{fig:vel_contour} exhibit higher velocity magnitude and more turbulent structures in the fridge when the fan operates at its maximum velocity.

Following the validation of our CHT solver against the experimental data, the main purpose of this study is to develop a data-driven ROM for the fridge for which a comprehensive database needs to be collected. The database must include a wider range of data for specific parameters, known as "sensitive parameters". These parameters tend to change the system non-linearly and the system's response to their modification is more amplified. Consequently, it is crucial to increase the sampling density of these parameters to ensure that the dataset is tailored and comprehensive for each specific problem.

The ambient temperature, evaporator temperature and fan angular velocity are the parameters of interest in this study.  To determine the most and least sensitive parameters, a sensitivity analysis was conducted for both fridge configurations by considering the most extreme values for each parameter: $T_{evap} = [-15, +4]\; ^{\circ} C$, $T_{amb} = [16, 32] \; ^{\circ} C$, and $\omega = [0, 191.6] \: \text{rad/sec}$. These values are chosen based on the experimental tests. We carried out the simulations for the configurations, $C_1$ to $C_8$, with combinations of parameters reported in \tab{tab:experimental_config}.

\begin{table}[h!]
\centering
 \begin{tabular}{||c |c |c |c |c |c |c |c |c||} 
 \hline
Configuration  & $C_1$ & $C_2$ & $C_3$ & $C_4$ & $C_5$ & $C_6$ & $C_7$ & $C_8$  \\ [0.5ex] 
 \hline\hline
 $T_{evap} \: (^{\circ} C)$ & -15 & -15 & -15 & -15& 4 & 4 & 4 & 4 \\ \hline
$T_{amb} \: (^{\circ} C)$  & 16& 16& 32& 32& 16& 16& 32& 32 \\ \hline
 $\omega \: (\%)$         & 0& 100& 0& 100& 0& 100& 0& 100 \\ \hline
 \end{tabular}
 \caption{Sensitivity analysis: simulation plan of various configurations to assess system sensitivity.}
 \label{tab:experimental_config}
\end{table}

\begin{figure}[t]
    \centering
        \includegraphics[width=100mm,scale=0.5]{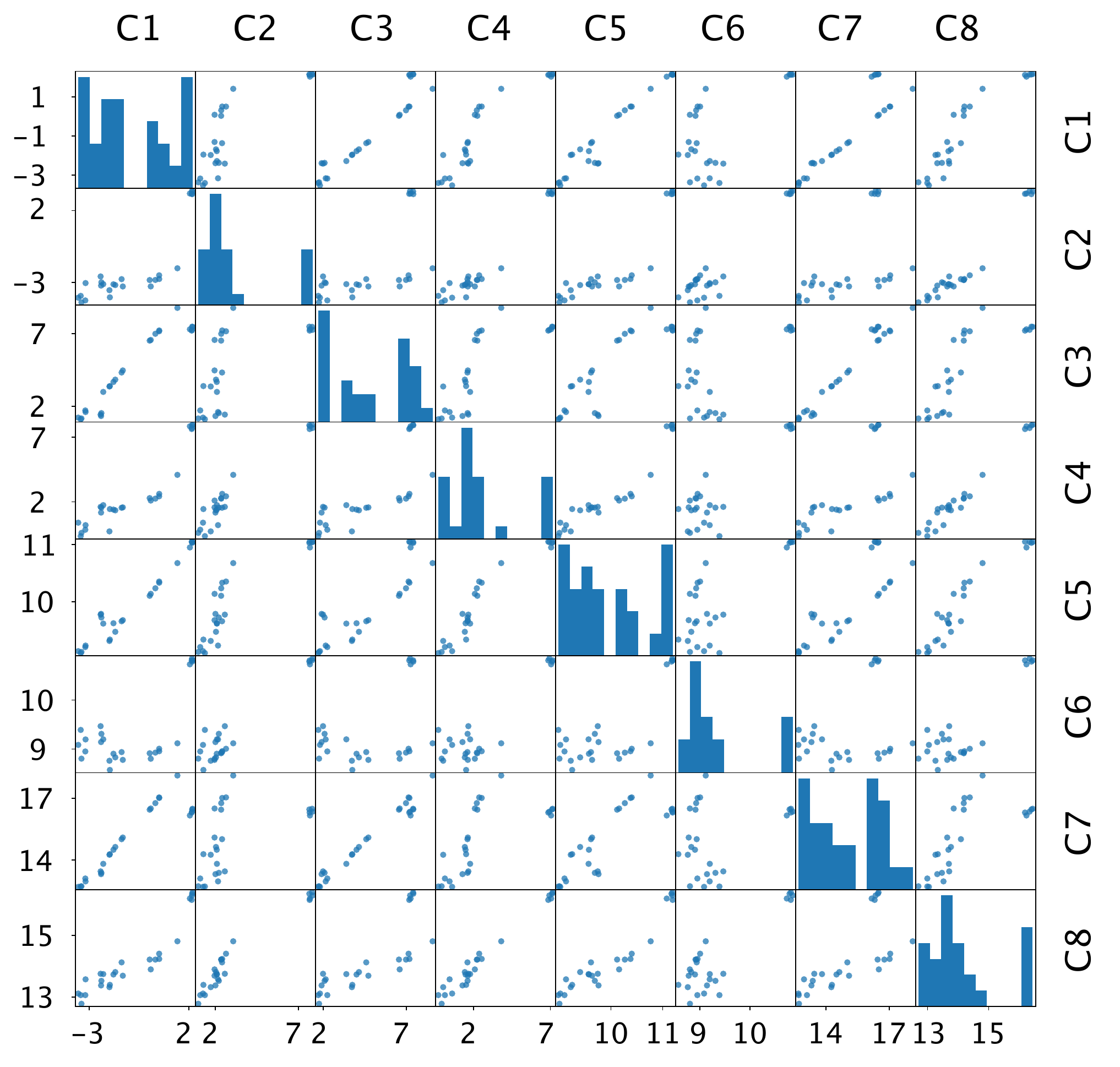}
        \caption{Sensitivity analysis: scatter matrix of temperature at sensor points for the configurations listed in \tab{tab:experimental_config}. The diagonal elements represent the variance of each configuration, while the off-diagonal ones indicate the covariance between the two configurations.}
        \label{fig:sensitivity analysis}
\end{figure}

To analyze the system sensitivity with respect to the parameters, we computed the scatter matrix of temperature at the sensor points for the configurations listed in \tab{tab:experimental_config}. The scatter matrix is a square matrix where the diagonal grids represent the variance of each configuration, while the off-diagonal grids indicate the covariance between pairs of configurations. Each point within the grids represents the relationship between a pair of temperatures at the same sensor. The distribution of the points within the grids defines the relationship among configurations. If the points follow a polynomial pattern of any order, there is a corresponding polynomial relationship between configurations. Conversely, the scattered points in a random pattern without any clear trend suggest a nonlinear relationship. 

The results of this analysis are displayed in \fig{fig:sensitivity analysis}. We noticed a significant non-linear relationship when the fan velocity varies from $0$ to $191.6$ (rad/sec), as indicated by the covariance between $C_1$ and $C_2$, as well as $C_1$ and $C_6$ configurations. On the other hand, changing the ambient and evaporator temperatures does not introduce non-linearity to the solution; the points mostly exhibit a linear correlation although some outliers are observed.      Based on this analysis, we realized that fan velocity is the most sensitive parameter in our problem which requires a denser sampling. Consequently, 11 sampling points for this parameter were considered, ranging from $\omega_{min}=0 \;$ to $\omega_{max}= 191.6 \;$ (rad/sec) with an increment of $10 \%$. For the remaining parameters, 4 evaporator temperatures:  $T_{evap} = [-15,-7.9,-3.25, +4]\; ^{\circ} C$ and 2 ambient temperatures: $T_{amb} = [16, 32] \; ^{\circ} C$ were considered.  The combination of these parameters results in a total of 88 various configurations. A CHT simulation has been conducted for each configuration and the results were collected to build the database.

%%%%%%%%%%%%%%%%%%%%%%%%%%%%%%%%%
%\newpage 

\section{\textbf{Surrogate Model}}\label{sec:Rom_res}
\begin{figure}[ht]
    \centering
        \includegraphics[width=60mm,scale=0.3]{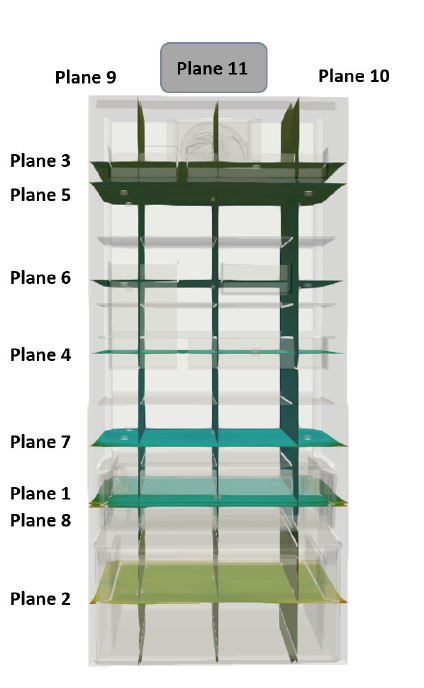}
        \caption{Fridge Planes of interest for the temperature field reconstruction.}
        \label{fig:planes}
\end{figure}

Considering the computational expense of the high-fidelity numerical simulation, we develop reduced-order models to reconstruct the full temperature field at an unknown parameter value, where limited information (such as measurements at the sensor locations) or no information is available. First, we demonstrate the application of the POD-RBF method in Section \ref{sec:res_PODI} to reconstruct the temperature field at an unknown parameter value based on the solutions of the temperature field available at known parameter values. Next, we reconstruct the temperature field from a few sensor locations' temperature data (it can be both experimental or numerical, however, we considered numerical results for demonstration) using the conventional Gappy POD approach in Section \ref{sec:res_GPOD}. Finally, a stable deep learning-enhanced GPOD approach named ANNGPOD is employed in Section \ref{sec:res_anngpod} and Section \ref{sec:res_anngpodexp} to reconstruct the full temperature field provided sparse experimental and/or numerical datasets only at a few sensor locations. Figure \ref{fig:planes} show 3 vertical and 8 horizontal planes where the temperature field is reconstructed. The horizontal planes are placed in such a way that they contain the sensor locations as shown in Fig. \ref{fig:sensor_loc}. To carry out the parametric study, the evaporator temperature $T_{ev}$, ambient temperature $T_{amb}$, and percentage of fan velocity, $v_f$ are considered. As indicated in the previous section \ref{sec:res}, from the sensitivity analysis, the training dataset density should be higher in the parametric direction of the fan velocity. Hence, the fan velocity ranges from $0$ to $100$ $\%$ of the full throttle which is $1600$ rpm in step of 10 $\%$ i.e., total of $11$, two ambient temperatures i.e., $16$ $^\circ C$ and $32$ $^\circ C$, and $4$ evaporator temperatures i.e., $-15$ $^\circ C$, $-7.9$ $^\circ C$, $-3.25$ $^\circ C$ and $4$ $^\circ C$ are considered. Therefore, a total of 88 parametric locations are considered to generate high-fidelity simulation data. The total spatial size of the dataset combining the grid points of all the selected planes is 1443297. Therefore, the size of the input matrix is $\mathbb{R}^{(88 \times 3)}$  whereas the size of the output matrix is $\mathbb{R}^{( 88 \times 1443297)}$ considering the total parametric locations, $88$, type of parameters, $3$ and total spatial location, $1443297$. Now, in the following subsections, we demonstrate several reduced-order models of interest taking a part of the input dataset as training points and the rest parameter values as validation points.

\subsection{POD-RBF ROM}\label{sec:res_PODI} 
\begin{figure}[ht]
    \centering \includegraphics[width=75mm,scale=0.3]{{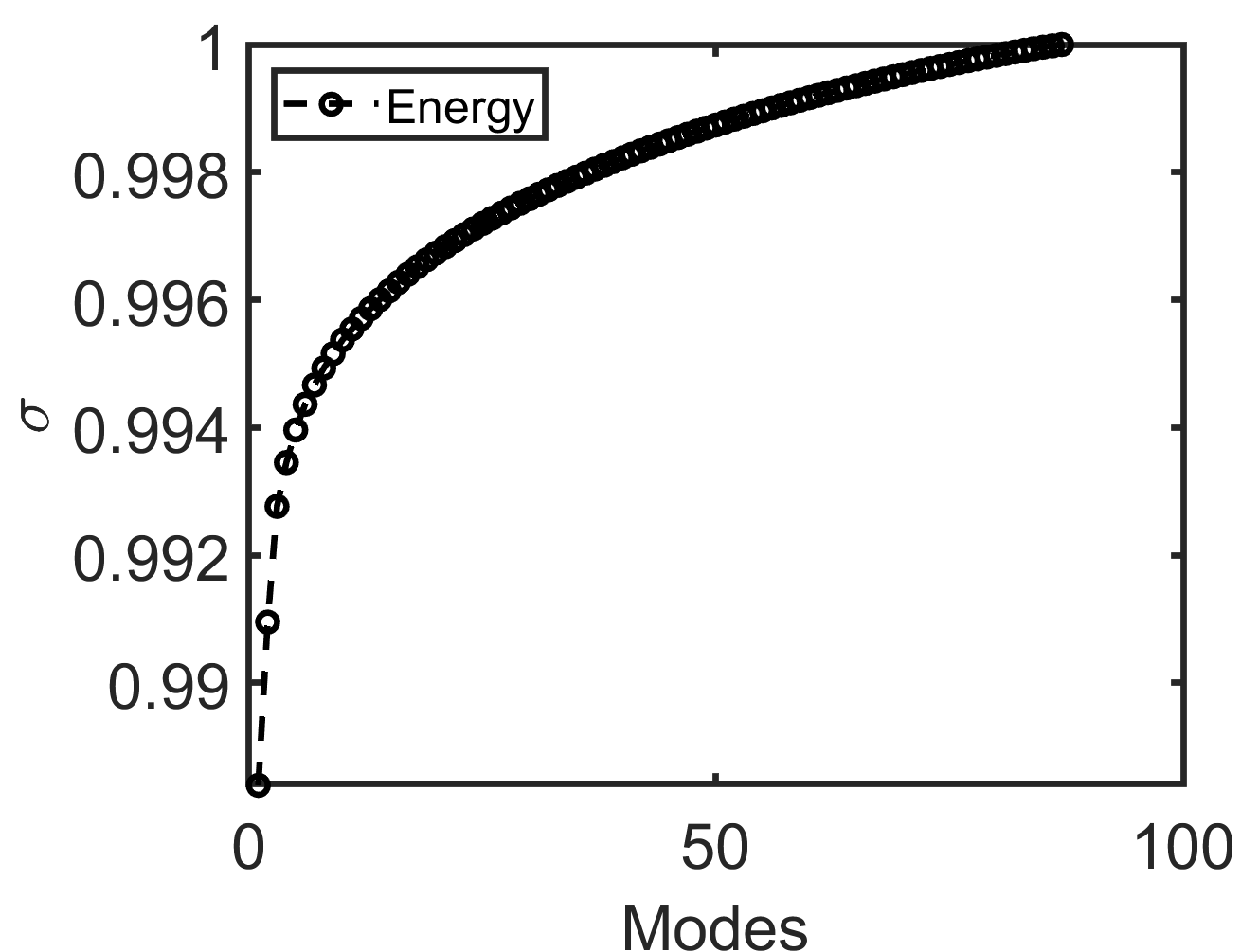}}
        \caption{Cumulative energy with different number of modes associated with temperature field.}
        \label{fig:Energy}
\end{figure}

 As mentioned earlier in the Section \ref{sec:method_PODI}, the POD-RBF ROM consists of two stages: POD-based projection followed by the approximation using RBF interpolation. To minimize the prediction error using this ROM technique, the projection error must first be minimized. Singular Value Decomposition (SVD) is first applied to the dataset containing 88 parametric locations to compute the POD modes. Figure \ref{fig:Energy} shows cumulative energy or eigenvalues associated with the number of POD modes as shown in Eqn.\ref{eq:energy_modes}. With $3$, $10$, $25$ and $58$ modes, the system contains 99$\%$, 99.5$\%$, 99.7$\%$ and 99.9$\%$ of energy respectively.   

\begin{figure}[ht]
\centering
% \hspace{-3cm}
\subfloat[ $\sigma$ = 0.99]{\label{fig:PODreconserror99}\includegraphics[width=.15\linewidth]{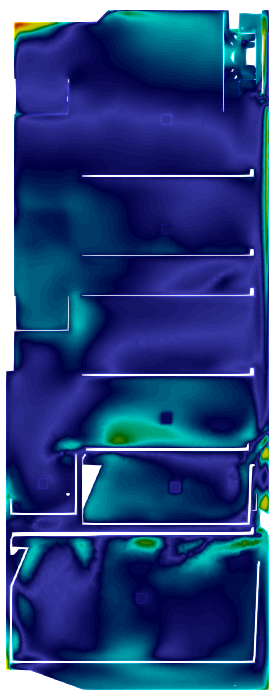}}
\subfloat[$\sigma$ = 0.995]{\label{fig:PODreconserror995}\includegraphics[width=.15\linewidth]{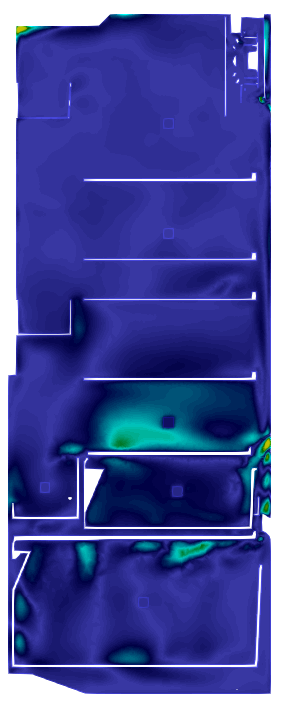}}
\subfloat[$\sigma$ = 0.997]{\label{fig:PODreconserror995}\includegraphics[width=.15\linewidth]{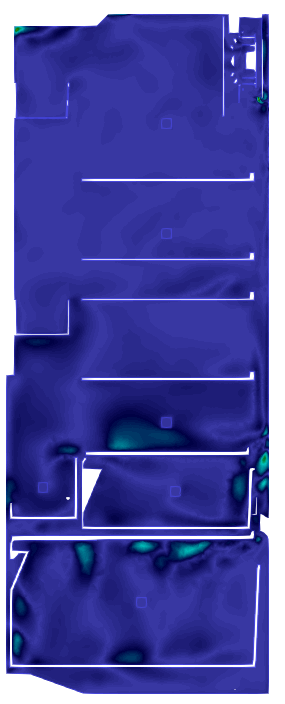}}
\subfloat[$\sigma$ = 0.999]{\label{fig:PODreconserror995}\includegraphics[width=.315\linewidth]{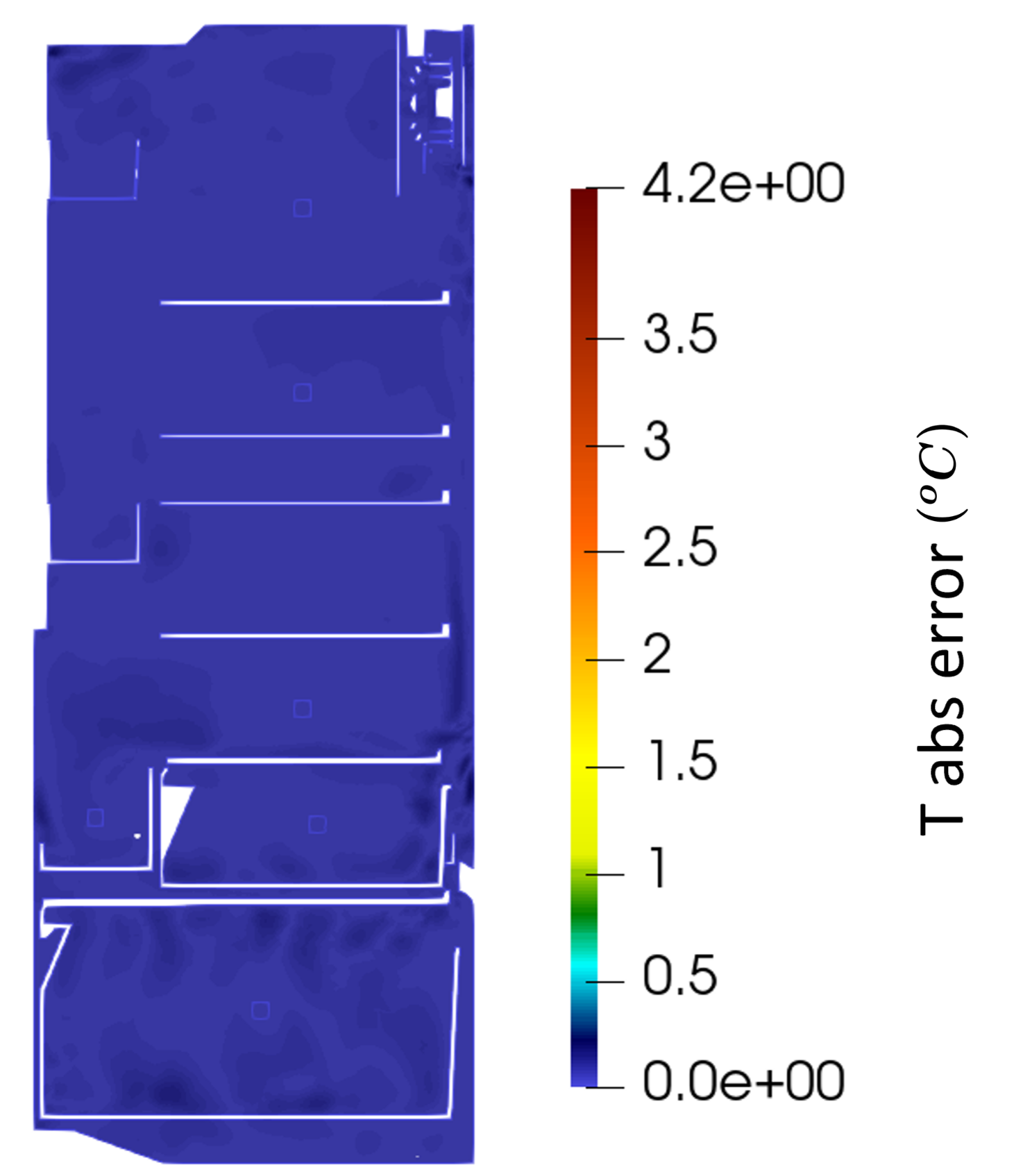}}
\caption{POD-based reconstruction error associated with the temperature field of the mid-section at $x=0$}
\label{fig:POD_Reconstruction}
\end{figure}

The temperature field is reconstructed at the mid-section, $x = 0$ considering different numbers of POD modes at a parametric point associated with $T_{ev}$ of -15 $^{\circ}C$, $T_{amb}$ of 16 $^{\circ}C$ and $v_f$ is kept at 100$\%$ as shown in Fig.\ref{fig:POD_Reconstruction}. At $v_f$ of 100$\%$, temperature field possesses significant non-linearity as compared to the condition when the fan velocity is less - therefore, this particular parametric point is chosen to assess the absolute error distribution when different numbers of POD modes are used for reconstruction. The absolute temperature error is computed as $\left|T_{C F D}-T_{R O M}\right|$, where the $T_{C F D}$ is the high-fidelity temperature field and the $T_{ROM}$ is the temperature field associated with the surrogate model computed at the vertex centres ($n$ = $1443297$) of the computational cells contained in the 11 planes of interest. It is noticed, at $\sigma = 0.99$, the maximum error related to it is 3.17 $^o C$. The maximum temperature error occurred at the top-left corner of the section due to possible stagnation points and absolute temperature rise. However, the non-linearity of the flow physics is increased inside the crisper (lowermost chamber of the fridge) due to the generation of a flow vortex, as the air inside the fridge gets swept away by the fan situated at the top of the fridge,  which eventually raises the temperature prediction error. However, at $\sigma = 0.995$, $\sigma = 0.997$, $\sigma = 0.999$ the maximum error associated with it is 1.68 $^o C$, 0.97 $^o C$ and 0.18 $^o C$. With $\sigma = 0.997$ and $\sigma = 0.999$, the projection-error distribution is less than 0.25 $K$ throughout the sections. The temperature error distribution is significantly low for the other 10 sections when the number of modes is considered more than 25 and $\sigma$ is higher or equal to $0.997$.

\begin{figure}[ht]
\centering
\begin{subfigure}[b]{0.75\textwidth}
\centering
% \hspace{-3cm}
\subfloat[ $x$ = 0.0]{\label{fig:RBFPOD_error_1}\includegraphics[width=.20\linewidth]{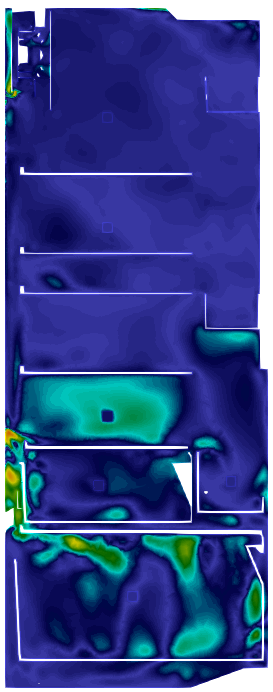}}
\subfloat[$x$ = 0.14725]{\label{fig:RBFPOD_error_2}\includegraphics[width=.20\linewidth]{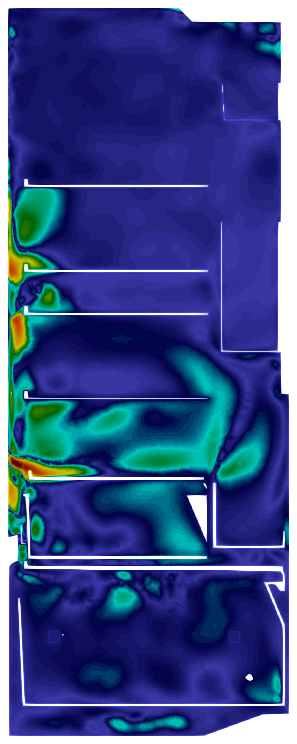}}
\subfloat[$x$ = -0.14725]{\label{fig:RBFPOD_error_3}\includegraphics[width=.385\linewidth]{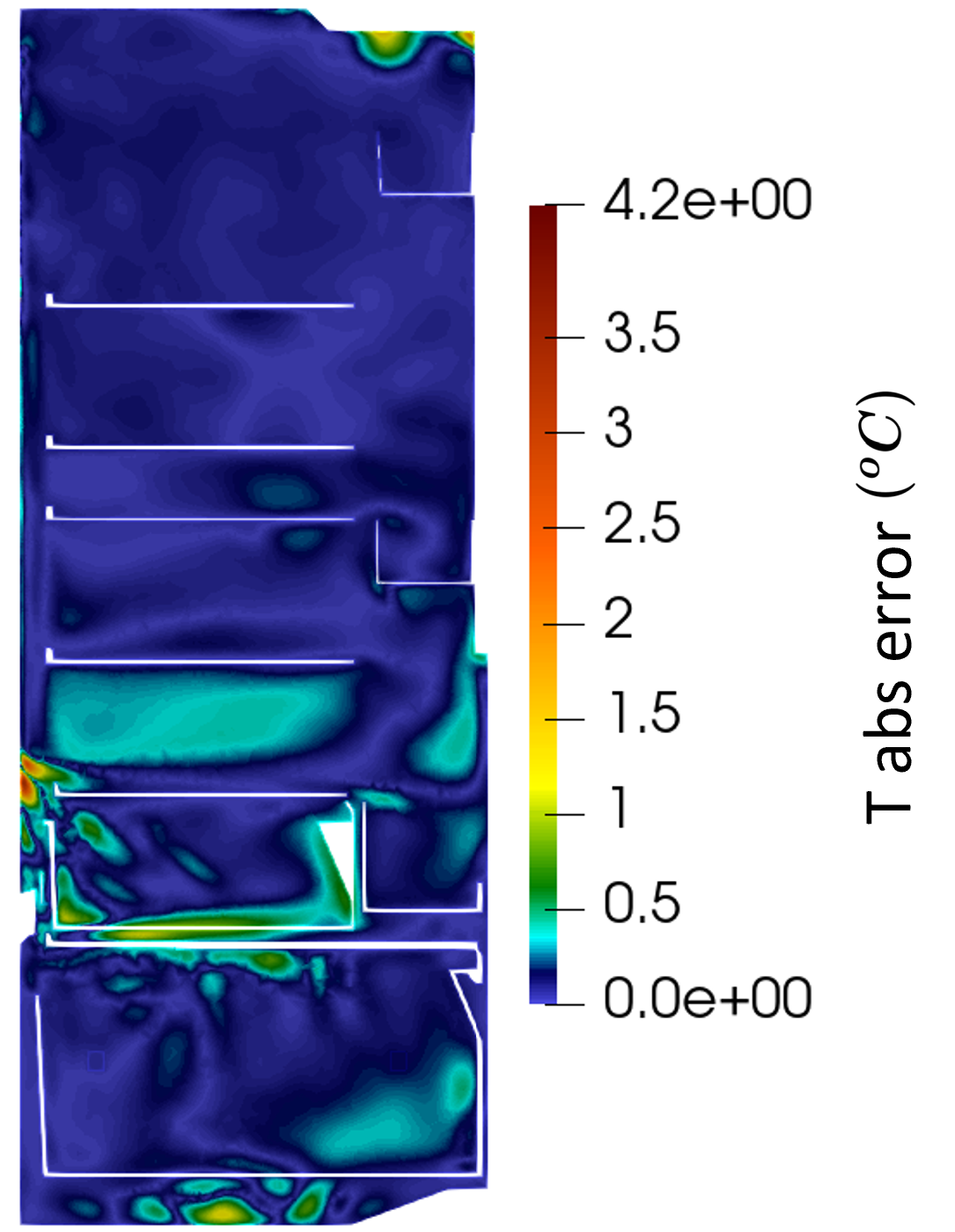}}
\end{subfigure}
\medskip
\begin{subfigure}[b]{0.75\textwidth}
% \hspace{-3cm}
\centering
\subfloat[ $z$ = 0.77795]{\label{fig:RBFPOD_error_4}\includegraphics[width=.20\linewidth]{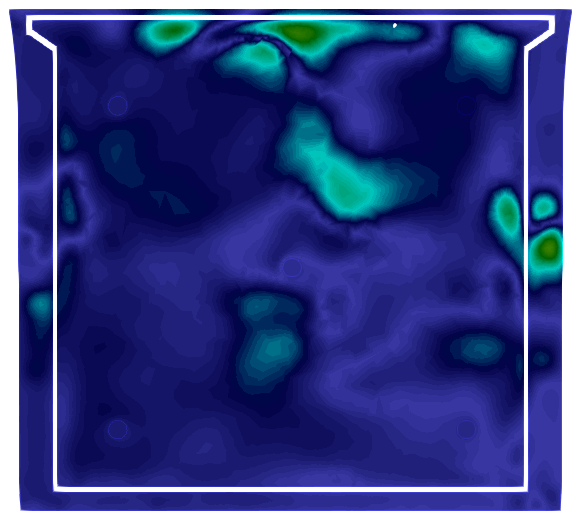}}
\subfloat[$z$ = 0.95944]{\label{fig:RBFPOD_error_5}\includegraphics[width=.20\linewidth]{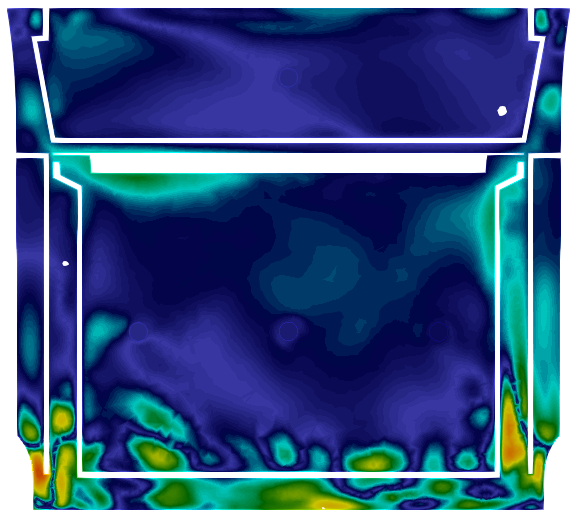}}
\subfloat[$z$ = 0.96615]{\label{fig:RBFPOD_error_6}\includegraphics[width=.20\linewidth]{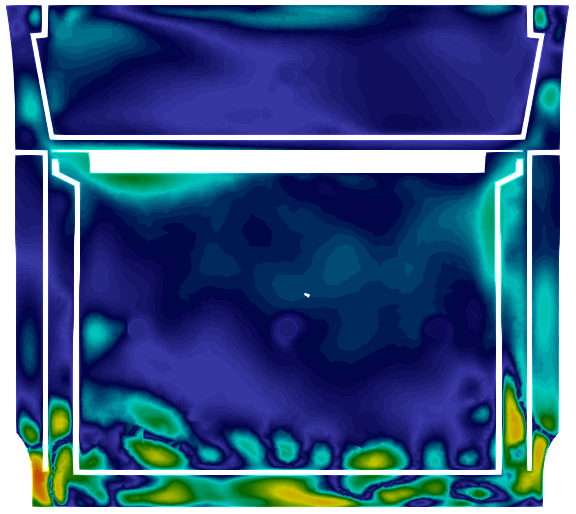}}
\subfloat[$z$ = 1.07215]{\label{fig:RBFPOD_error_7}\includegraphics[width=.20\linewidth]{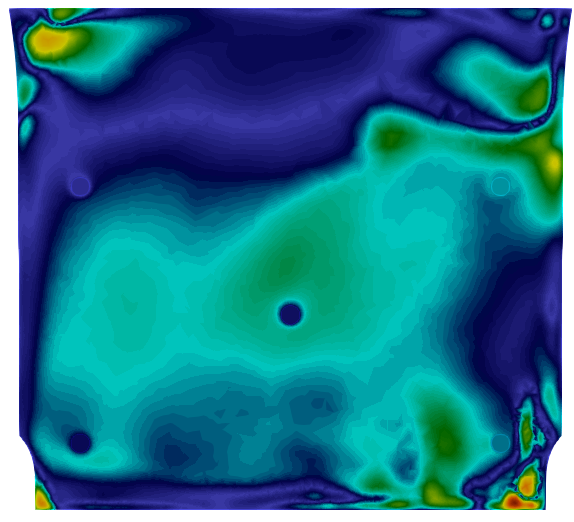}}
\end{subfigure}
\medskip
\begin{subfigure}[b]{0.75\textwidth}
\centering
% \hspace{-3cm}
\subfloat[ $z$ = 1.24475]{\label{fig:RBFPOD_error_8}\includegraphics[width=.20\linewidth]{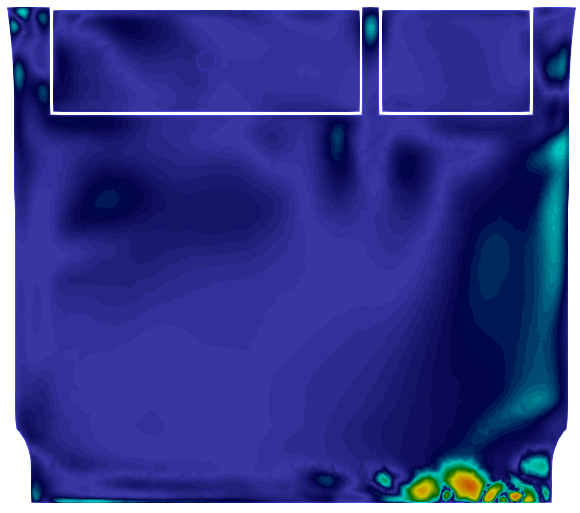}}
\subfloat[$z$ = 1.38115]{\label{fig:RBFPOD_error_9}\includegraphics[width=.20\linewidth]{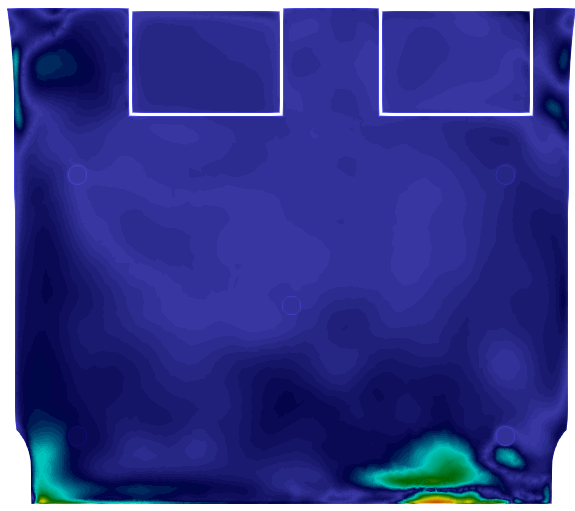}}
\subfloat[$z$ = 1.56115]{\label{fig:RBFPOD_error_10}\includegraphics[width=.20\linewidth]{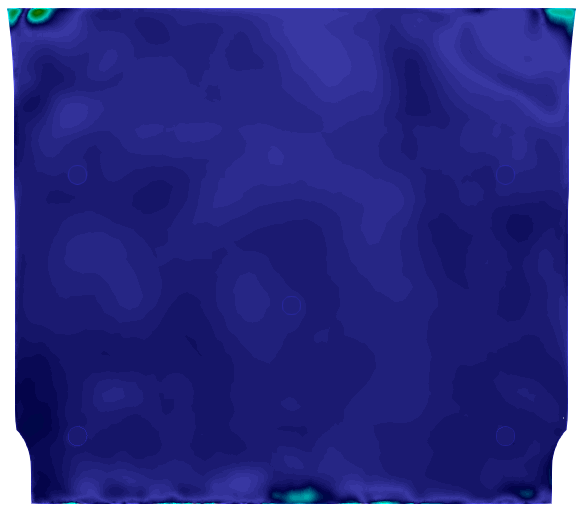}}
\subfloat[$z$ = 1.59975]{\label{fig:RBFPOD_error_11}\includegraphics[width=.20\linewidth]{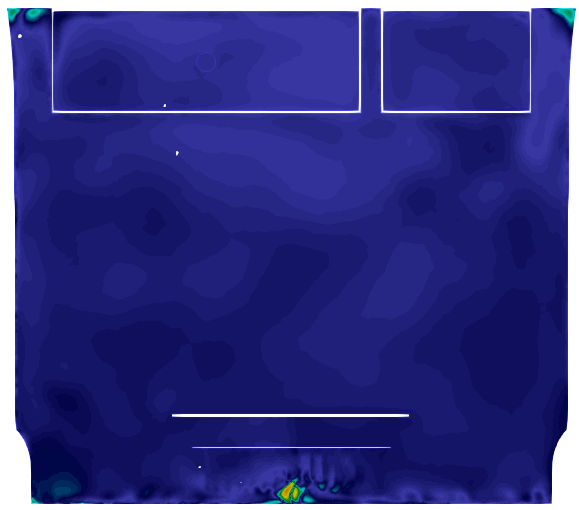}}
\end{subfigure}
\caption{Reconstruction error associated with the temperature field using POD-RBF method}
\label{fig:PODI_error}
\end{figure}

Now, to employ POD-RBF ROM, We aim to reconstruct the temperature field corresponding to a particular parameter value while the rest 87 parametric points are considered to be training points, the reduced coefficient obtained from the temperature fields associated with the training parameter values are interpolated using the RBF approach to reconstruct the temperature field associated with an unknown parameter of interest. Fig.\ref{fig:PODI_error}, shows the absolute temperature error distribution at different planes corresponding to $T_{amb} = 16 ^\circ C$, $T_{ev} = -15 ^\circ C$, and $f_v = 100 \%$. However other than a few locations such as the start of any narrow passages, where the velocity drops due to stagnation, temperature increases thereby increasing the absolute temperature error as well, the majority of the spatial locations show an error distribution less than $0.5$ $^o C$.

\begin{figure}[ht]
\centering
\begin{subfigure}[b]{0.88\textwidth}
\centering
% \hspace{-3cm}
\subfloat[ normal scale]{\label{fig:rbfpodmeanerror_normal}\includegraphics[width=.50\linewidth]{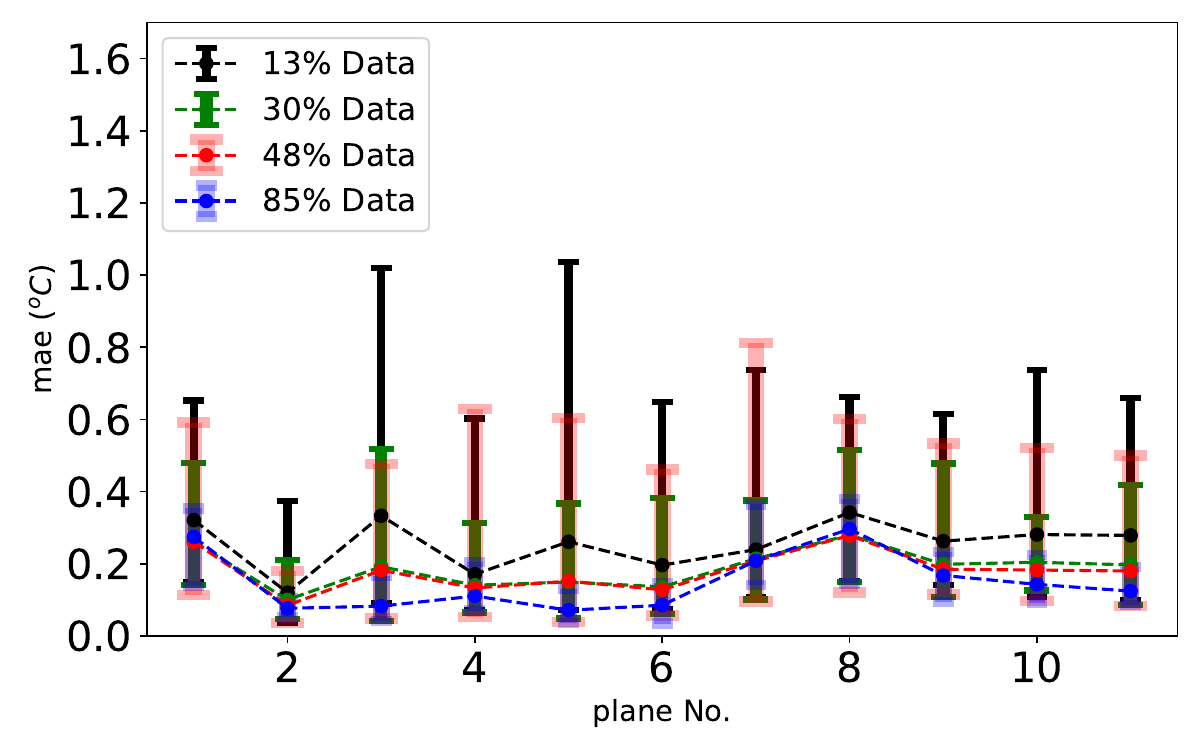}}
\subfloat[log scale]{\label{fig:rbfpodmeanerror_log}\includegraphics[width=.50\linewidth]{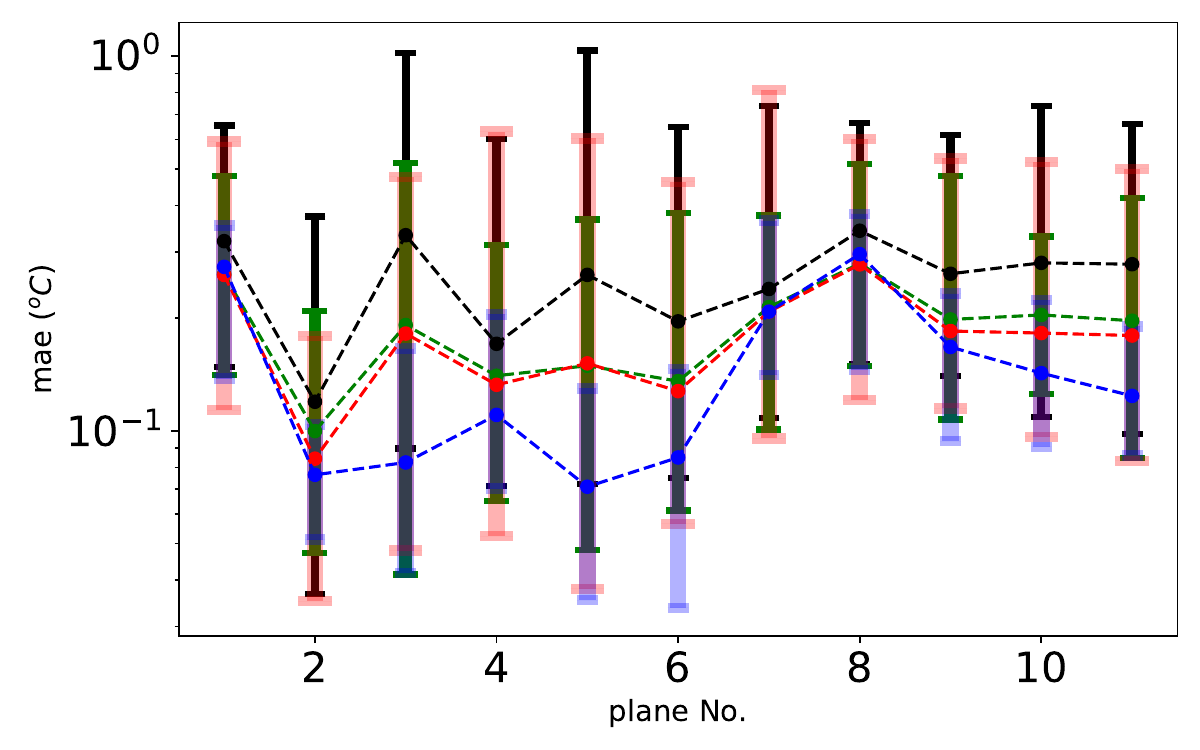}}
\end{subfigure}
\medskip
\caption{Mean absolute error associated with the prediction of POD-RBF ROM with benchmark CFD data using different percentages of training datasets.}
\label{fig:meanerror_diffdata}
\end{figure}

Next, considering the computationally expensive full-order simulation, we aim to reduce the computation effort required to generate the dictionary of the training dataset by sparsifying the number of training points to produce accurate ROM. Previously in Fig.\ref{fig:PODI_error}, the size of training parametric points is 87 ($~ 99 \%$ of the total input size), but now we will consider $85 \%$, $48 \%$ , $30 \%$ and $13 \%$ of the total input size. Figure \ref{fig:meanerror_diffdata} shows prediction error associated with different percentages of data. The parameter values outside the training datasets are considered as the validation points. Therefore, $15 \%$, $52 \%$, $70 \%$ and $87 \%$ corresponding to the above-mentioned training sets are available for validation. The average ($\text{MAE}_\text{avg}$) and the maximum ($\text{MAE}_\text{max}$) over all the validation points' mean absolute error (MAE) are computed as shown in Eqn \ref{rbfpod1}: 

\begin{equation}
\label{rbfpod1}
\begin{aligned}
& \text{MAE}_\text{avg}=\frac{1}{n_{val}}\left(\sum_{n_{val}}\left(\frac{1}{n_{planes}} \sum_{n_{planes}}\left|\left(T_{C F D}-T_{ROM}\right)\right|\right)\right) \\
&\text{MAE}_\text{max}= \max_{n_{val}}\left(\frac{1}{n_{planes}} \sum_{n_{planes}}\left|\left(T_{C F D}-T_{R O M}\right)\right|\right)
\end{aligned}
\end{equation}

where $n_{planes}$ and $n_{val}$ are the total number of planes of interest inside the fridge and validation points. The $\text{MAE}_\text{avg}$ and $\text{MAE}_\text{max}$ associated with the validation points are shown in Fig.\ref{fig:meanerror_diffdata}. The dotted line indicates the $\text{MAE}_\text{avg}$ whereas the upper bound of the graph shows the $\text{MAE}_\text{max}$ corresponding to different percentages of the training datasets considered. It is evident from \ref{fig:rbfpodmeanerror_normal} and \ref{fig:rbfpodmeanerror_log}, when the training data size is large which is $85 \%$, $\text{MAE}_\text{avg}$ is minimum at most of the planes except for planes no. $1$, $7$, $8$. However the improvement of the $\text{MAE}_\text{avg}$ and $\text{MAE}_\text{max}$ is not linear with the enrichment of the training dataset. When the training dataset size is $13$ $\%$, the $\text{MAE}_\text{avg}$ is maximum at all the planes. The $\text{MAE}_\text{max}$, similarly shows minimum prediction accuracy at $13$ $\%$ of the training data, whereas maximum prediction accuracy at the most dense dataset considering the majority of the planes. The $\text{MAE}_\text{avg}$ for all the training sets lies around $0.25$ $^o C$ whereas the $\text{MAE}_\text{max}$ does not cross the error limit of $1$ $^o C$ even at most sparse training dataset.      

% \GS{The introduction is still missing but we should clearly write the overall story behind the paper. For example, why did we switch from POD-RBF to POD-ANN? RH: This point was Addressed after the ANNGPOD methodology, check whether it is satisfactory or not.}

%%%%% GS did not comment after this %%%%%%%%%%%%%%%%%%%%%%%%%%%%%%%%%%%%%%%%%

 \subsection{Gappy POD}\label{sec:res_GPOD} 
\begin{figure}[ht]
\centering
\begin{subfigure}[b]{0.75\textwidth}
\centering
% \hspace{-3cm}
\subfloat[ $x$ = 0.0]{\label{fig:GPODerror1_5sensor}\includegraphics[width=.20\linewidth]{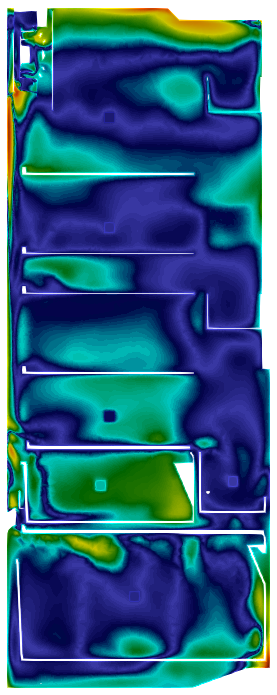}}
\subfloat[$x$ = 0.14725]{\label{fig:GPODerror2_5sensor}\includegraphics[width=.20\linewidth]{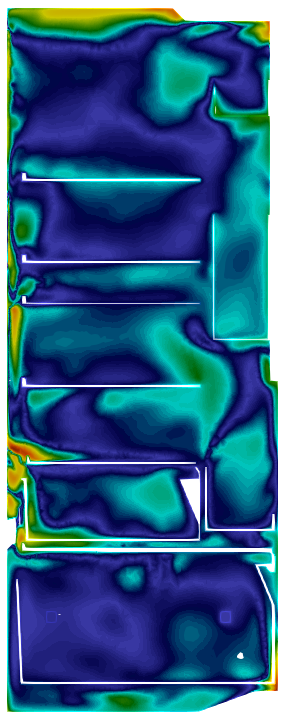}}
\subfloat[$x$ = -0.14725]{\label{fig:GPODerror3_5sensor}\includegraphics[width=.385\linewidth]{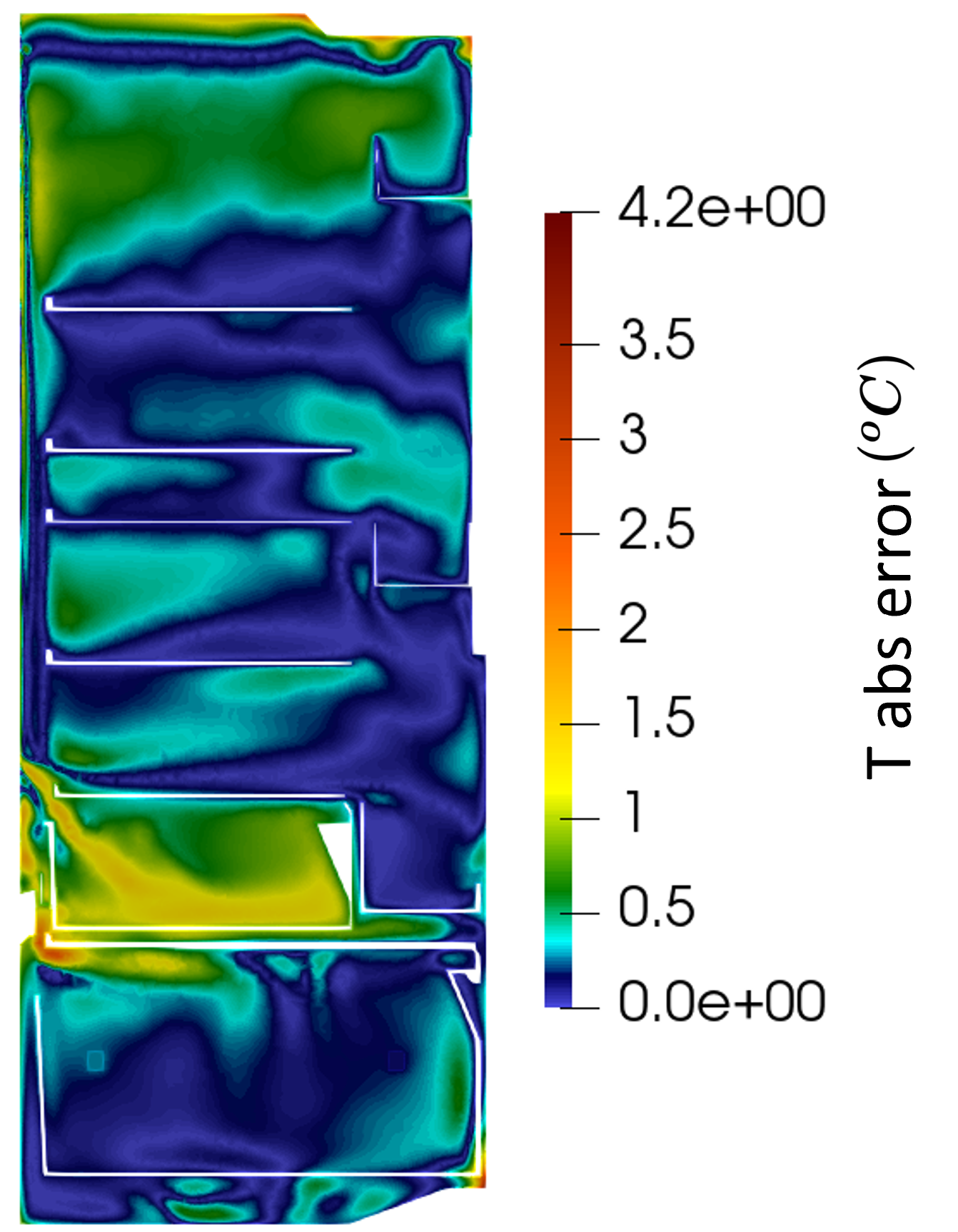}}
\end{subfigure}
\medskip
\begin{subfigure}[b]{0.75\textwidth}
% \hspace{-3cm}
\centering
\subfloat[ $z$ = 0.77795]{\label{fig:GPODerror4_5sensor}\includegraphics[width=.20\linewidth]{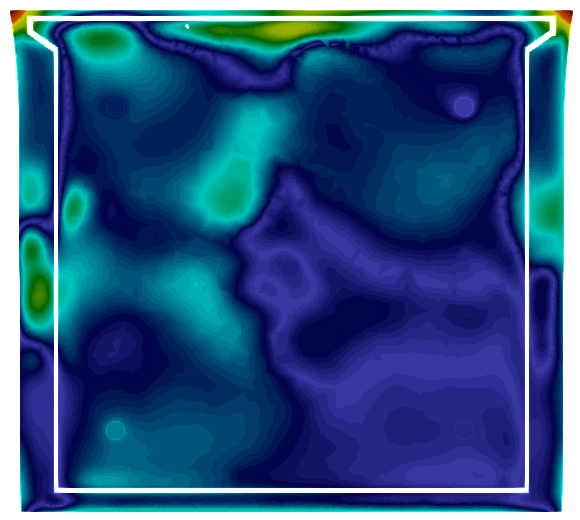}}
\subfloat[$z$ = 0.95944]{\label{fig:GPODerror5_5sensor}\includegraphics[width=.20\linewidth]{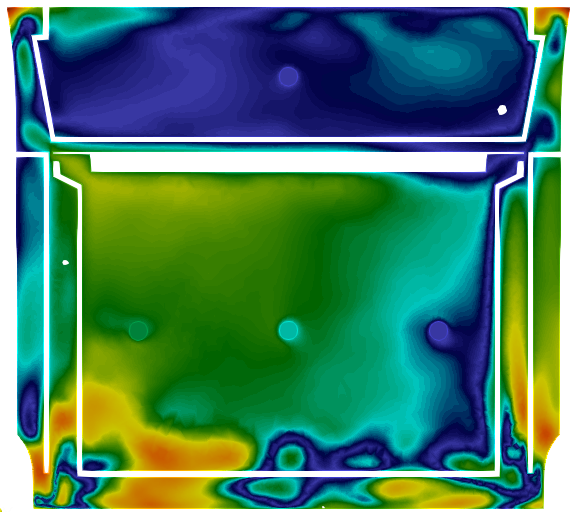}}
\subfloat[$z$ = 0.96615]{\label{fig:GPODerror6_5sensor}\includegraphics[width=.20\linewidth]{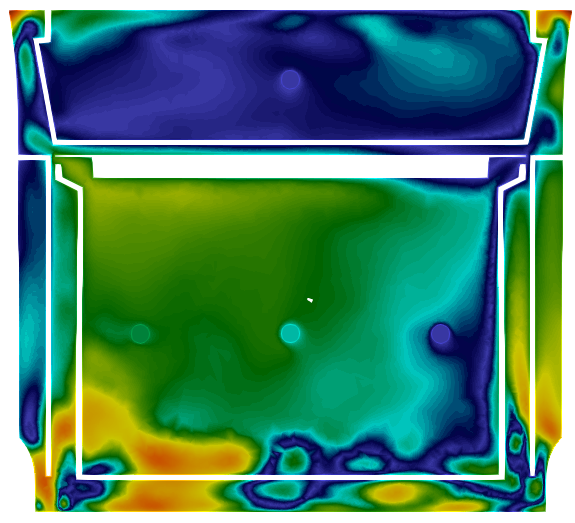}}
\subfloat[$z$ = 1.07215]{\label{fig:GPODerror7_5sensor}\includegraphics[width=.20\linewidth]{img/GPOD_error_5_sensors/sec096615.png}}
\end{subfigure}

\medskip
\begin{subfigure}[b]{0.75\textwidth}
\centering
% \hspace{-3cm}
\subfloat[ $z$ = 1.24475]{\label{fig:GPODerror8_5sensor}\includegraphics[width=.20\linewidth]{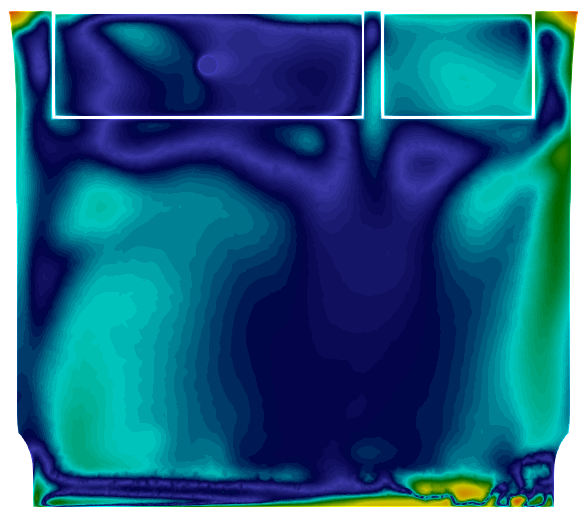}}
\subfloat[$z$ = 1.38115]{\label{fig:GPODerror9_5sensor}\includegraphics[width=.20\linewidth]{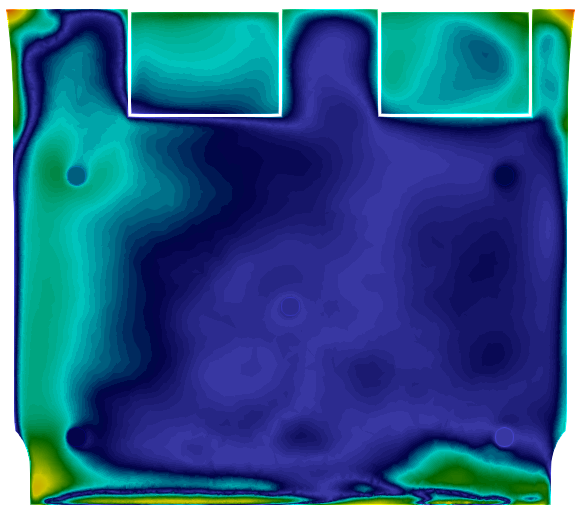}}
\subfloat[$z$ = 1.56115]{\label{fig:GPODerror10_5sensor}\includegraphics[width=.20\linewidth]{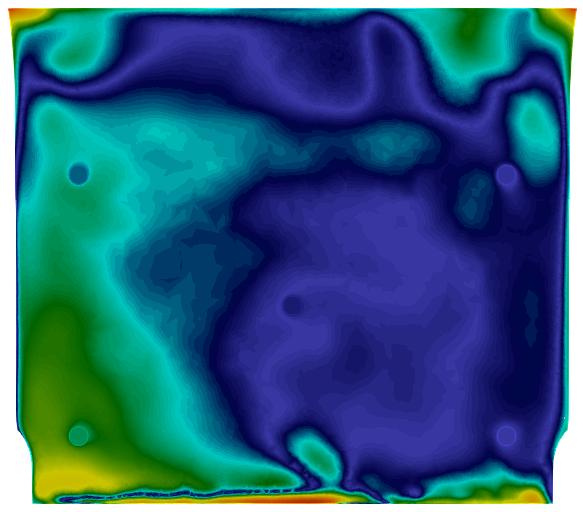}}
\subfloat[$z$ = 1.59975]{\label{fig:GPODerror11_5sensor}\includegraphics[width=.20\linewidth]{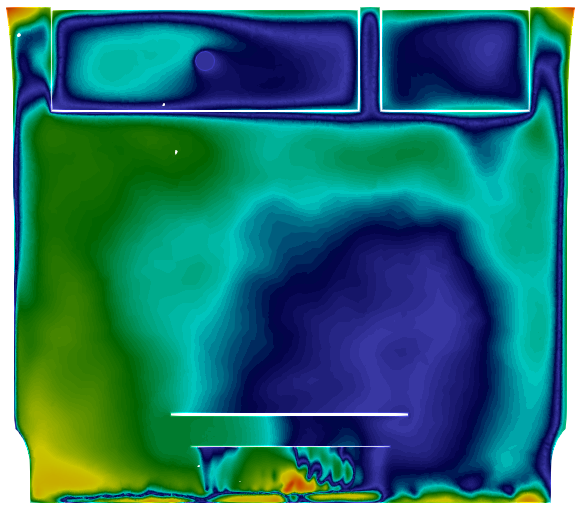}}
\end{subfigure}

\caption{Reconstruction error associated with the temperature field using Gappy POD approach.}
\label{fig:GPOD_error}
\end{figure}

In this current section, the temperature field is reconstructed using the temperature measurements at a few sensor locations \cite{willcox2006unsteady}. As discussed in section \ref{sec:method_GPOD}, for the application of Gappy POD, apart from the training datasets where high-fidelity results are available, a sparse temperature dataset is also available at all or a few locations of the $26$ sensors shown in \fig{fig:sensor_loc} corresponding to validation parameter values. If the number of sensors and the number of POD modes are different, $\Phi_R$ matrix in Eqn. \ref{eq:GPOD4} is rectangular. To avoid the computation of the pseudo-inverse ($\dag$) of the truncated $\Phi$ matrix $(\Phi_R)$ and additional prediction error as a result of it, we have maintained the number of modes equal to the number of sensor locations. Furthermore, the prediction accuracy associated with the conventional Gappy POD is very sensitive to the location and number of the sensors. In this section, we will show the absolute and mean temperature field reconstruction error associated with different planes of interest considering only the $5$ mid-sensor locations as indicated by $1.3$, $2.3$, $3.3$, $tp1$ and $cr3$ in \fig{fig:sensor_loc}. Figure \ref{fig:GPOD_error} shows that using the Gappy-POD approach, the error associated with the 11 planes is higher compared to the POD-RBF approach. Here, we have considered the parametric location corresponding to $T_{amb} = 16 ^\circ C$, $T_{ev} = -15 ^\circ C$, and $f_v = 100 \%$ for the validation and the rest $87$ parameters are training points. At sections, $z = 0.95944$, $z = 0.96615$ and $z = 1.07215$ the absolute error distribution crosses the error limit of $1$ $^o C$ at majority of the spatial locations. In the following section \ref{sec:res_anngpod}, we will show that with different numbers of sensor locations, the performance of the conventional GPOD further deteriorates, whereas our proposed ANNGPOD outperforms the classical method.

Similar to the previous section \ref{sec:res_PODI}, we now aim to reduce the training points to minimise the computational effort associated with the training dataset generation. Although, the $\text{MAE}_\text{avg}$ associated with $30 \%$, $48 \%$ and $85 \%$ training data is around $0.5$ $^o C$, the $\text{MAE}_\text{max}$ crosses the error limit of $1$ $^o C$, when the training points are sparsified beyond $48 \%$. Additionally, we observe that the GPOD ROM with $48 \%$ dataset shows poor performances as compared to the one with even sparse datasets, i.e., $30 \%$ training datasets at most of the planes.         
 
 \begin{figure}[ht]
\centering
\begin{subfigure}[b]{0.88\textwidth}
\centering
% \hspace{-3cm}
\subfloat[normal scale]{\label{fig:GPOD_5sensorsn}\includegraphics[width=.50\linewidth]{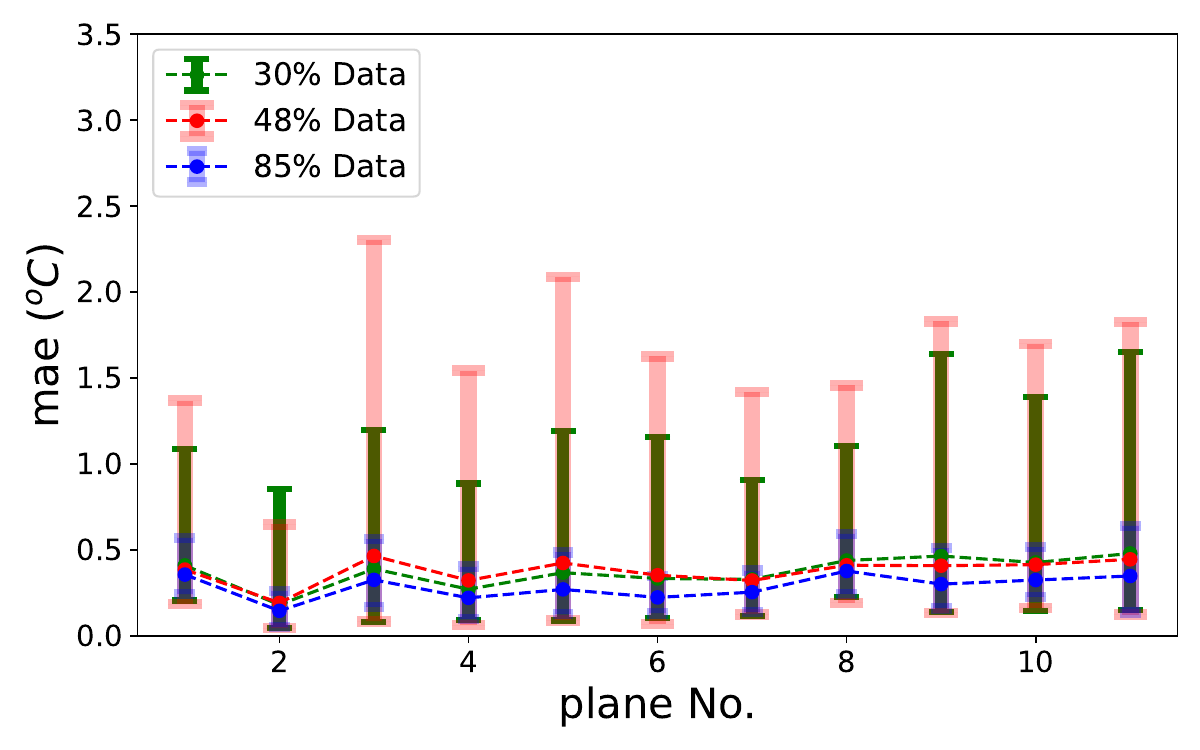}}
\subfloat[log scale]{\label{fig:GPOD_5sensorsl}\includegraphics[width=.50\linewidth]{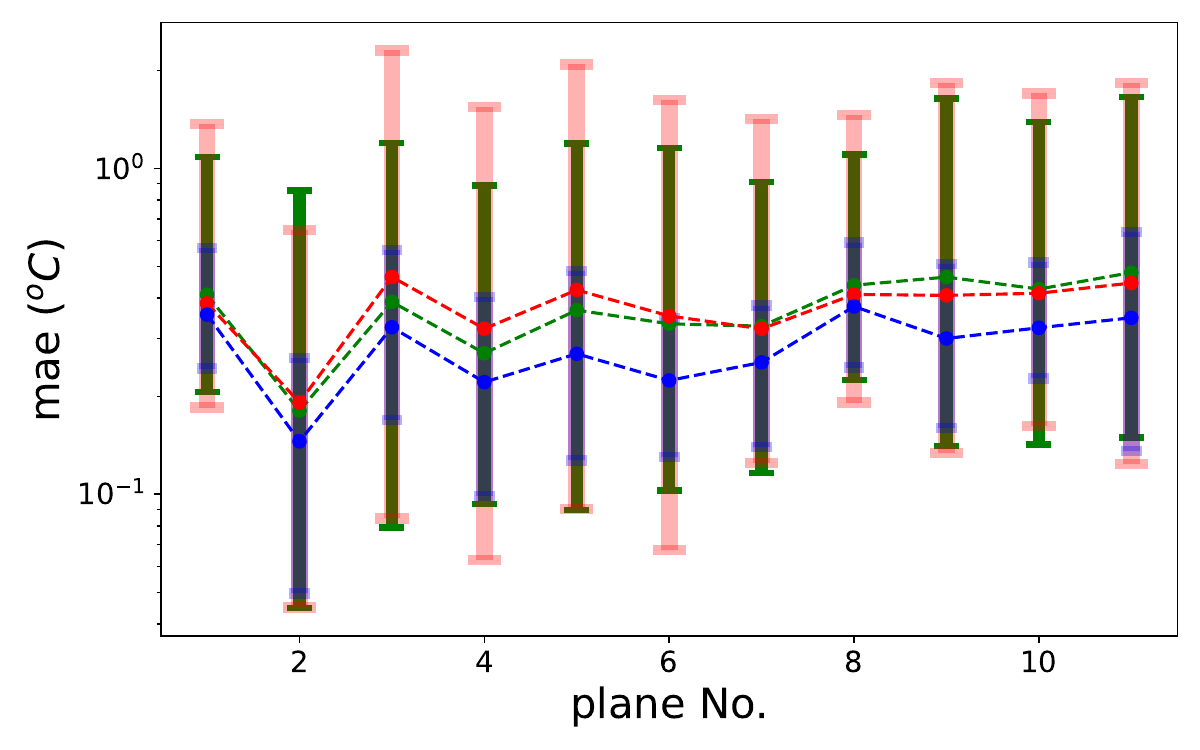}}
\end{subfigure}
\caption{Mean absolute error associated with the prediction of Gappy POD-based ROM with benchmark CFD data using different percentages of training datasets.}
\label{fig:GPODerror_5sensors_para}
\end{figure}

\subsection{Deep-learning enhanced Gappy POD} \label{sec:res_anngpod}

\begin{figure}[ht]
\centering
\begin{subfigure}[b]{0.75\textwidth}
\centering
% \hspace{-3cm}
\subfloat[ $p$ = 10, normal scale]{\label{fig:GPODANNGPOD_w1w21_10n}\includegraphics[width=.50\linewidth]{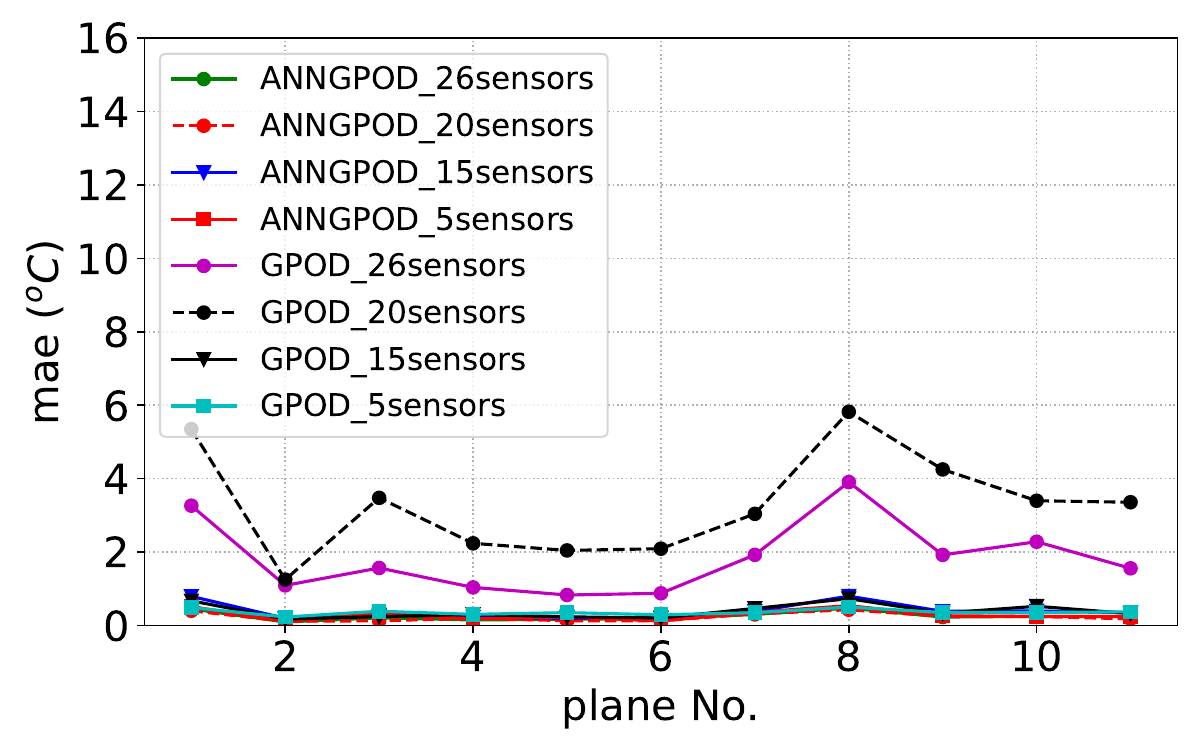}}
\subfloat[$p$ = 10, log scale]{\label{fig:GPODANNGPOD_w1w21_10l}\includegraphics[width=.50\linewidth]{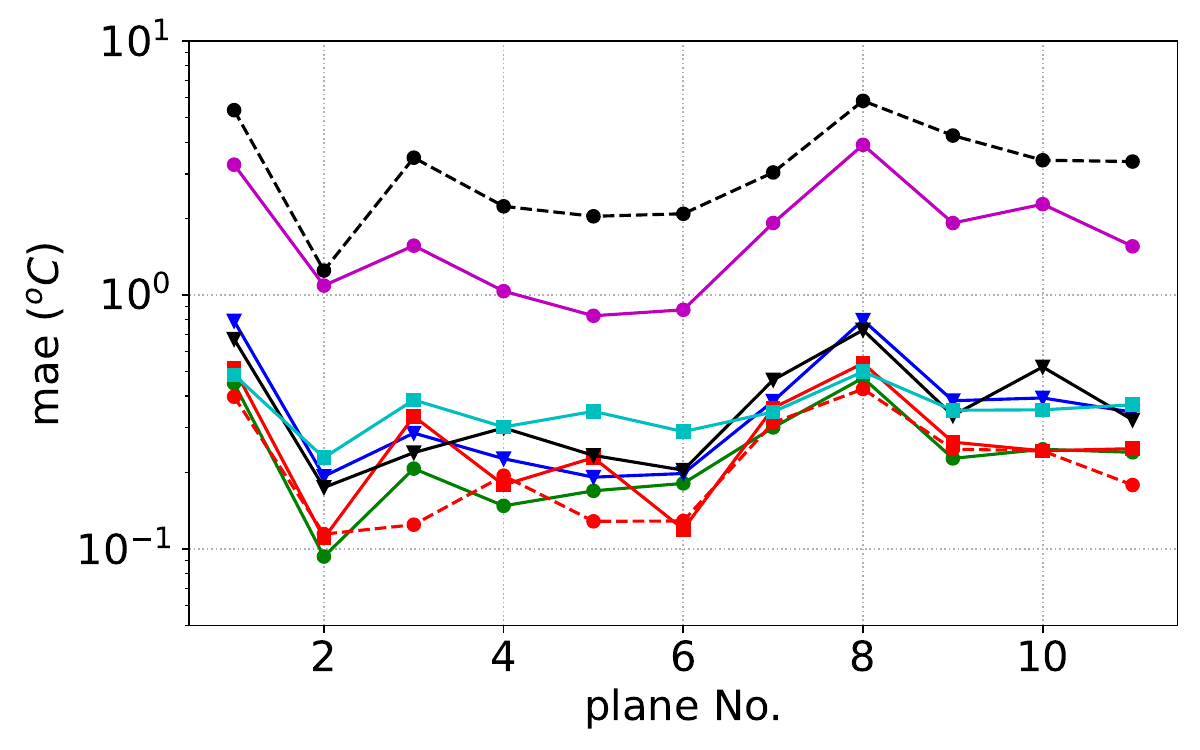}}
\end{subfigure}
\medskip
\begin{subfigure}[b]{0.75\textwidth}
\centering
% \hspace{-3cm}
\subfloat[ $p$ = 22, normal scale]{\label{fig:GPODANNGPOD_w1w21_22n}\includegraphics[width=.50\linewidth]{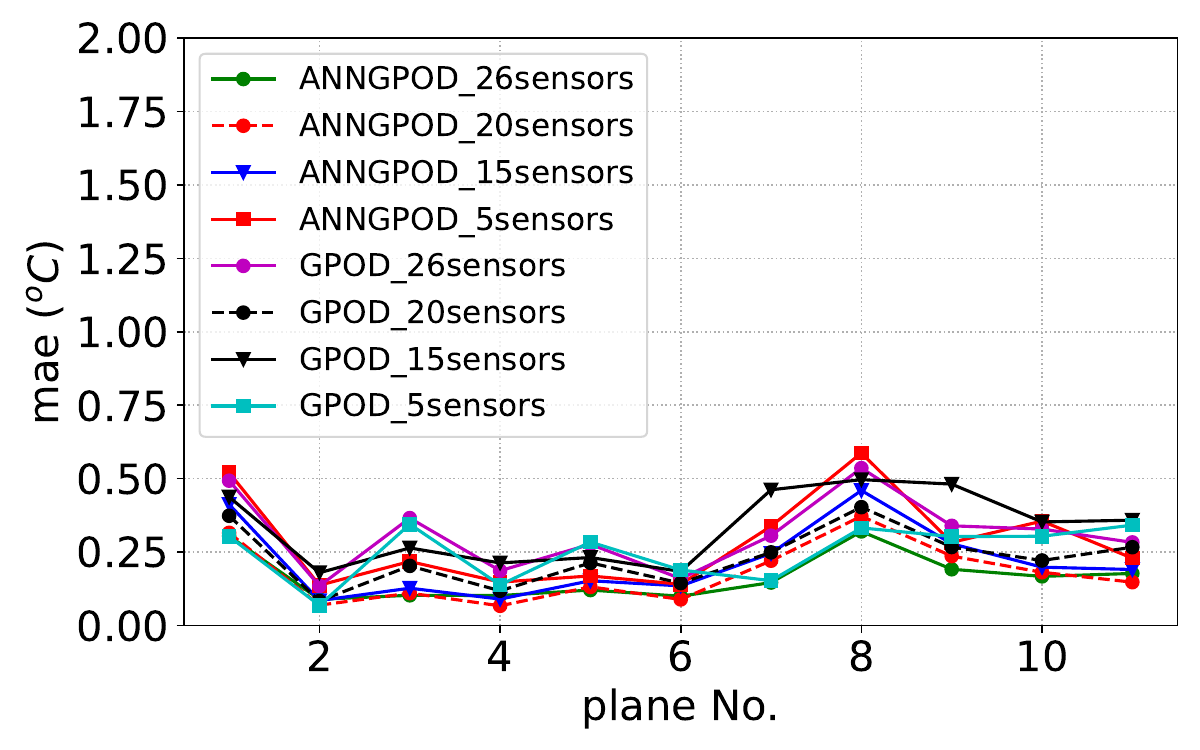}}
\subfloat[$p$ = 22, log scale]{\label{fig:GPODANNGPOD_w1w21_22l}\includegraphics[width=.50\linewidth]{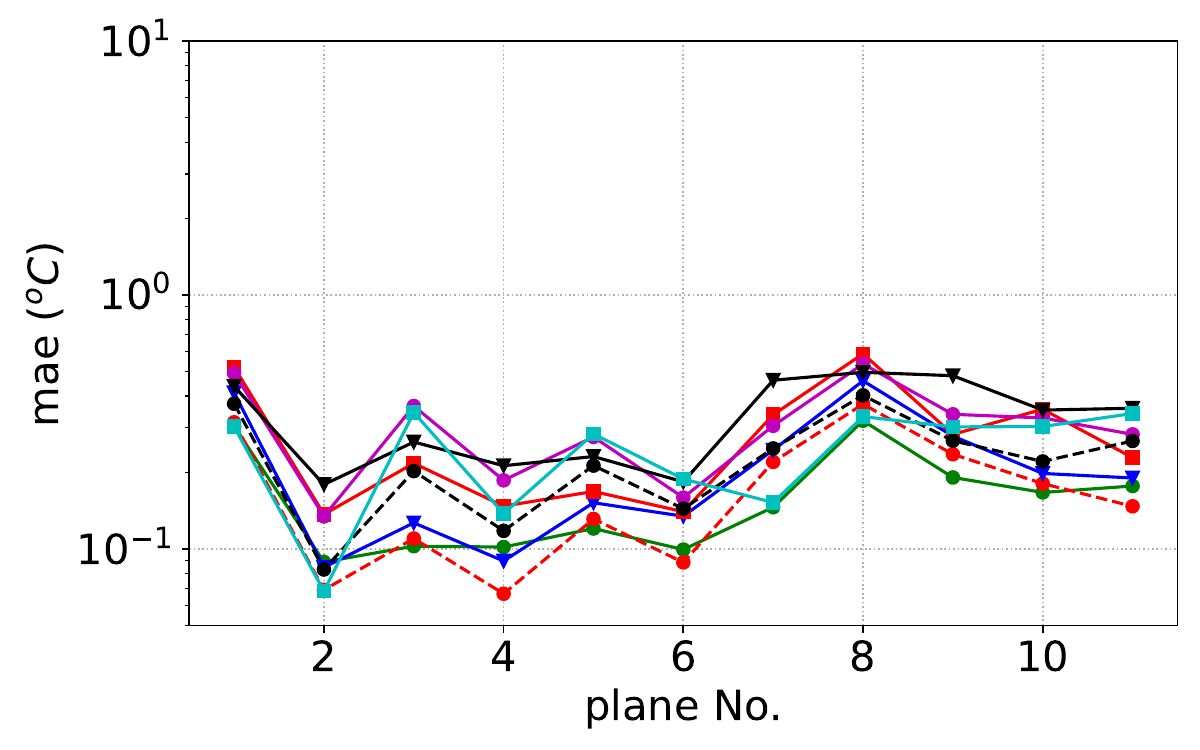}}
\end{subfigure}
\medskip
\begin{subfigure}[b]{0.75\textwidth}
\centering
% \hspace{-3cm}
\subfloat[ $p$ = 62, normal scale]{\label{fig:GPODANNGPOD_w1w21_62n}\includegraphics[width=.50\linewidth]{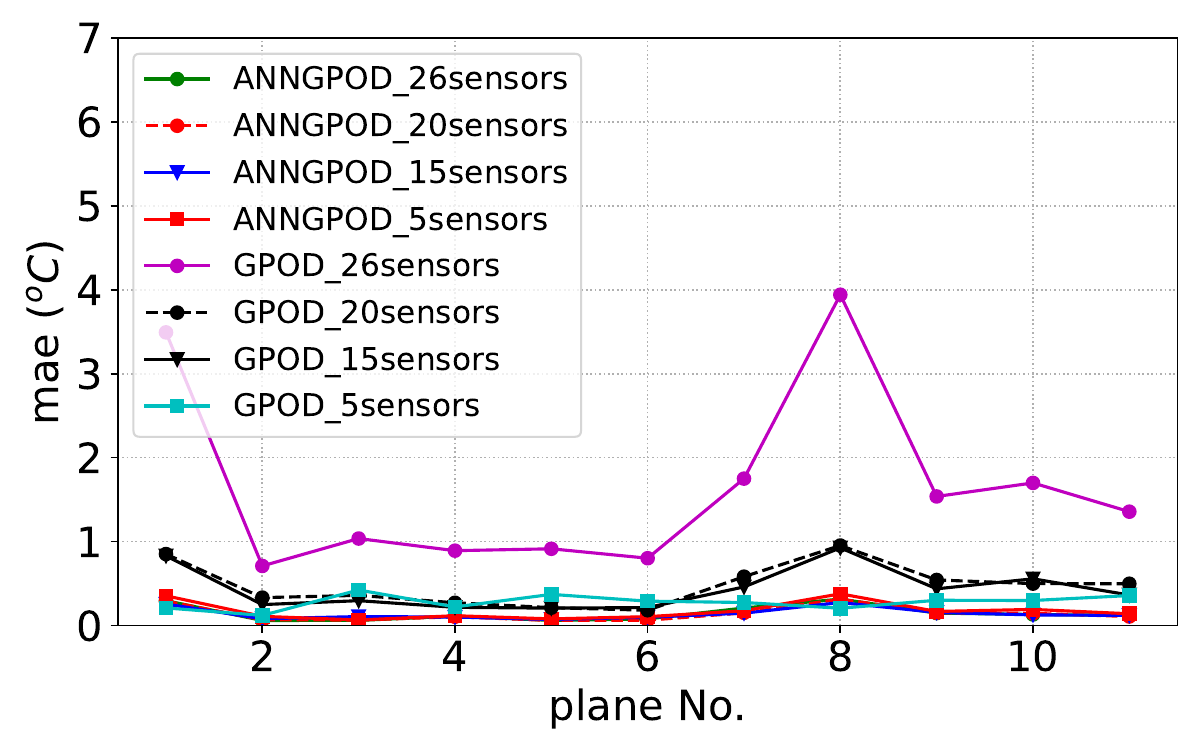}}
\subfloat[$p$ = 62, log scale]{\label{fig:GPODANNGPOD_w1w21_62l}\includegraphics[width=.50\linewidth]{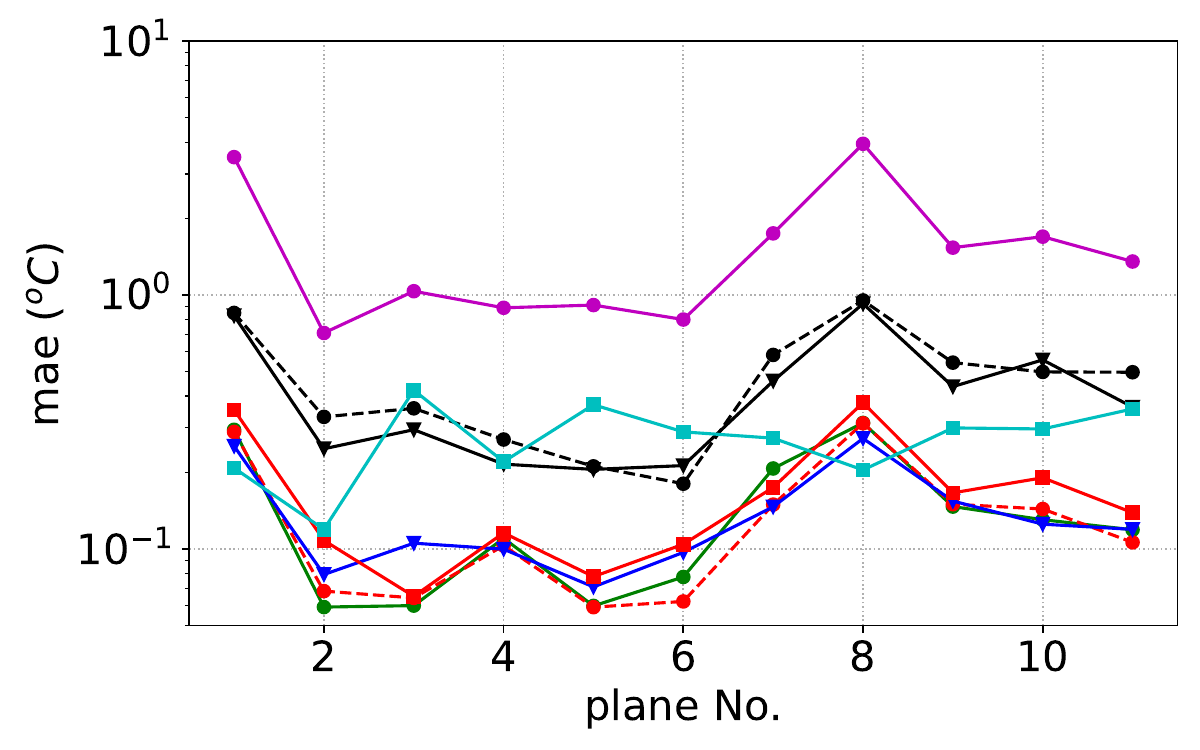}}
\end{subfigure}
\caption{Mean absolute error of temperature field prediction associated with different planes of interest using Gappy POD and ANNGPOD approach.}
\label{fig:GPODANNGPOD_error_w1w21}
\end{figure}

In this subsection, the deep-learning enhanced gappy-POD approach as shown in section \ref{sec:method_ANNGPOD} is demonstrated for the temperature field reconstruction at parameter values corresponding to validation points. In the case of the GPOD approach, the number of modes is kept equal to the number of sensor locations. Conversely, there are no similar constraints on the ANNGPOD method, since  Eqn. \ref{eq:anngpod} is solved using an optimization framework in the ANN network. Therefore the number of modes is decided in the ANNGPOD method based on how much energy of the dynamical system, the reduced order system must contain, such as $99.7$ $\%$ in the present case scenario. The number of sensor locations can be chosen independently of the number of modes considered. \fig{fig:GPODANNGPOD_error_w1w21} compares the MAE using GPOD and ANNGPOD methods while considering $87$ training points and $1$  validation point. Sparse temperature solutions at a few sensor locations are available at the validation parametric point. $4$ different numbers of sensor locations i.e., $5$, $15$, $20$ and $26$ are considered. In our current assessment of the mean absolute projection error at different planes, we will mainly focus on 3 different parameters varying the $T_{amb}$, $T_{ev}$, and $v_f$ i.e., $p = 10$ corresponds to $T_{amb} = 16 ^\circ C$, $T_{ev} = -15 ^\circ C$, and $v_f = 100 \%$;  $p = 22$ corresponds to $T_{amb} = 16 ^\circ C$, $T_{ev} = -7.9 ^\circ C$, and $v_f = 0 \%$ and $p = 62$ corresponds to $T_{amb} = 32 ^\circ C$, $T_{ev} = -3.25 ^\circ C$, and $v_f = 70 \%$. In \fig{fig:GPODANNGPOD_w1w21_10n} and \fig{fig:GPODANNGPOD_w1w21_10l}, $p=10$, $v_f$ of $100$ $\%$ introduces strong non-linearity in the flowfield. With $26$ and $20$ sensors, the prediction accuracy associated with the conventional GPOD method is poor as the MAE is above $1$ $^o C$ at all planes. Conversely, with the ANNGPOD method, the $MAE$ is within the acceptable limit for all the number of sensor locations. In the previous sub-section \ref{sec:res_GPOD}, the application of the GPOD approach in absolute error distribution is demonstrated with $5$ sensor locations. Form \fig{fig:GPODANNGPOD_w1w21_10l}, it is observed that prediction accuracy with $5$ sensors using the ANNGPOD method is better than the GPOD method with $5$ sensors at most of the planes except for the plane $7$. Although ANNGPOD does not show linear improvement in the prediction accuracy with the increase in the number of sensors, with sensors' numbers of $26$ and $20$, the $MAE$ error is much less than the ones associated with sensors' no. of $5$ and $15$. Conversely, the conventional GPOD does not show any correlation between the prediction accuracy with the number of sensors. It justifies the stability of our new approach ANNGPOD as compared to the classical GPOD method. In \fig{fig:GPODANNGPOD_w1w21_62n} and \fig{fig:GPODANNGPOD_w1w21_62l}, nonlinear convection is reduced as the $v_f$ is $70 \%$. GPOD with $26$ sensors is the only outlier among all the sensors' numbers variations using GPOD and ANNGPOD methods. On the contrary, in \fig{fig:GPODANNGPOD_w1w21_22n} and \fig{fig:GPODANNGPOD_w1w21_22l}, $v_f$ is $0$, that means the non-linearity of the flow physics is greatly reduced and only contributed by natural convection term. Here, both the GPOD and ANNGPOD method produces prediction errors within an acceptable limit of $1$ $^o C$. However, with $p = 22$ and $p = 62$, the ANNGPOD prediction with $20$ and $26$ sensors is better than the other variations of sensor numbers similar to what was observed in the case of $p = 10$. For the development of the ANNGPOD method, the POD-ANN approach is considered. In the ANN, $2$ intermediate hidden layers with $200$ and $64$ neurons are considered. The activation functions are $ReLU$, $Tanh$ and $Tanh$ associated with the hidden layers and the output respectively. A learning rate of $0.01$ with a decay rate of $0.75$ in step-size of $1000$ is considered. The number of epochs is chosen as $20000$. 

\begin{figure}[ht]
\centering
\begin{subfigure}[b]{1\textwidth}
\centering
% \hspace{-3cm}
\subfloat[ANNGPOD]{\label{fig:ANNGPOD_MID}\includegraphics[width=.20\linewidth]{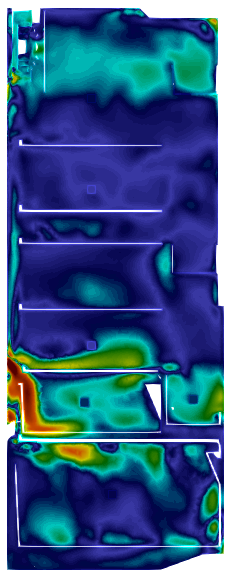}}
\subfloat[GPOD]{\label{fig:GPOD_MID}\includegraphics[width=.20\linewidth]{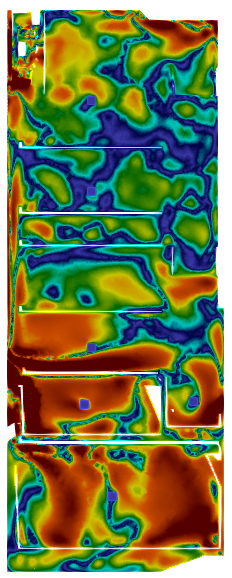}}
{\label{fig:scale}\includegraphics[width=.15\linewidth]{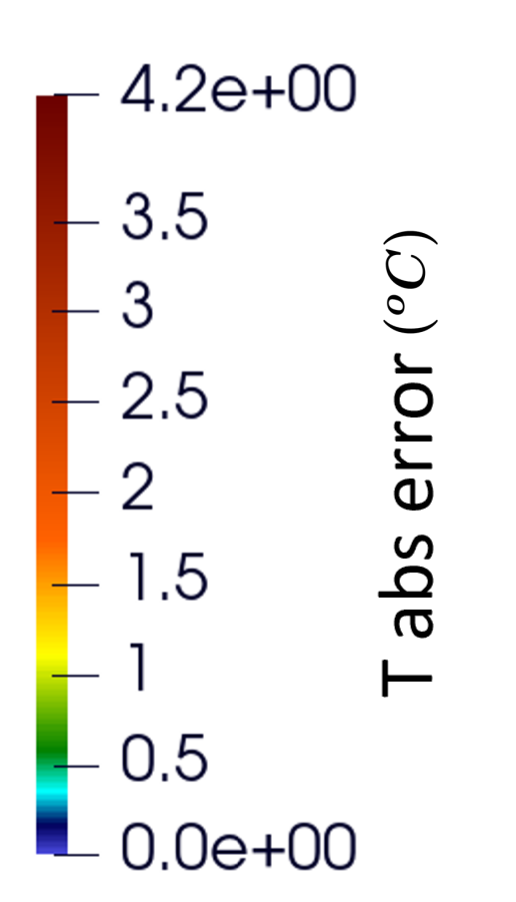}}
\end{subfigure}
\caption{Absolute error considering ANNGPOD and conventional Gappy POD with 26 sensors at the (x = 0) plane.}
\label{fig:GPODANNGPODerror_26sensors_para}
\end{figure}

\begin{figure}[ht]
\centering
\begin{subfigure}[b]{0.85\textwidth}
\centering
% \hspace{-3cm}
\subfloat[ $x$ = 0.0]{\label{fig:ANNGPODerror1_26sensor}\includegraphics[width=.20\linewidth]{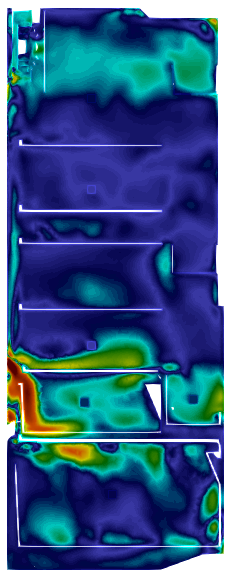}}
\subfloat[$x$ = 0.14725]{\label{fig:ANNGPODerror2_26sensor}\includegraphics[width=.20\linewidth]{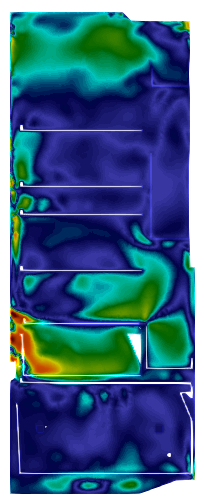}}
\subfloat[$x$ = -0.14725]{\label{fig:ANNGPODerror3_26sensor}\includegraphics[width=.19\linewidth]{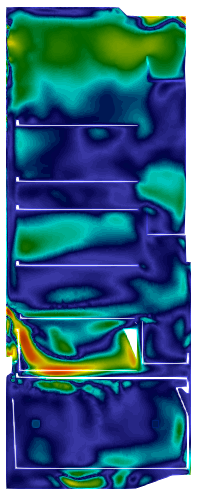}}
{\label{fig:scale_anngpod}\includegraphics[width=.21\linewidth]{img/ANNGPOD_error_26_sensors/scale1.png}}
\end{subfigure}
\medskip
\begin{subfigure}[b]{0.85\textwidth}
% \hspace{-3cm}
\centering
\subfloat[ $z$ = 0.77795]{\label{fig:ANNGPODerror4_26sensor}\includegraphics[width=.20\linewidth]{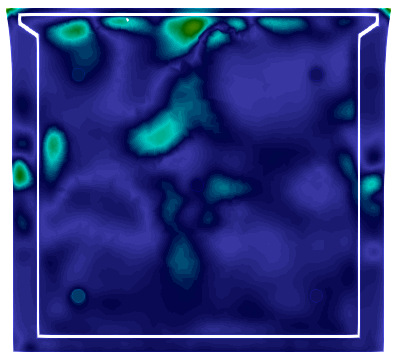}}
\subfloat[$z$ = 0.95944]{\label{fig:ANNGPODerror5_26sensor}\includegraphics[width=.20\linewidth]{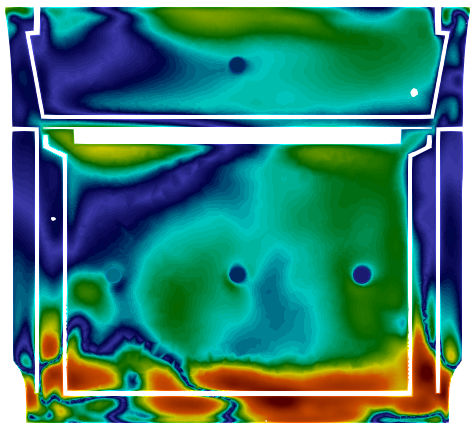}}
\subfloat[$z$ = 0.96615]{\label{fig:ANNGPODerror6_26sensor}\includegraphics[width=.20\linewidth]{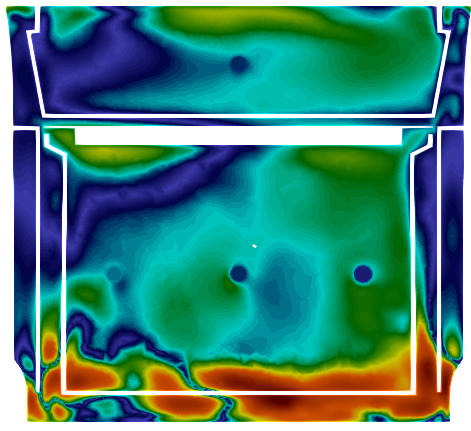}}
\subfloat[$z$ = 1.07215]{\label{fig:ANNGPODerror7_26sensor}\includegraphics[width=.20\linewidth]{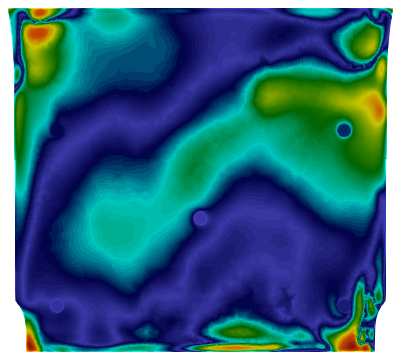}}
\end{subfigure}

\medskip
\begin{subfigure}[b]{0.85\textwidth}
\centering
% \hspace{-3cm}
\subfloat[ $z$ = 1.24475]{\label{fig:GPODerror8_5sensor}\includegraphics[width=.20\linewidth]{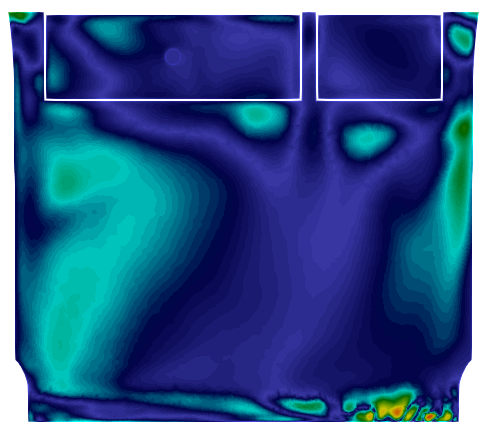}}
\subfloat[$z$ = 1.38115]{\label{fig:GPODerror9_5sensor}\includegraphics[width=.20\linewidth]{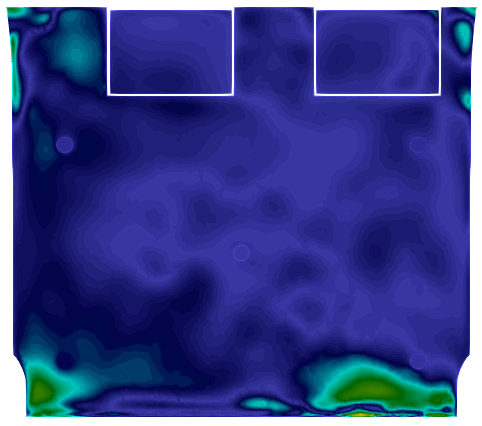}}
\subfloat[$z$ = 1.56115]{\label{fig:GPODerror10_5sensor}\includegraphics[width=.20\linewidth]{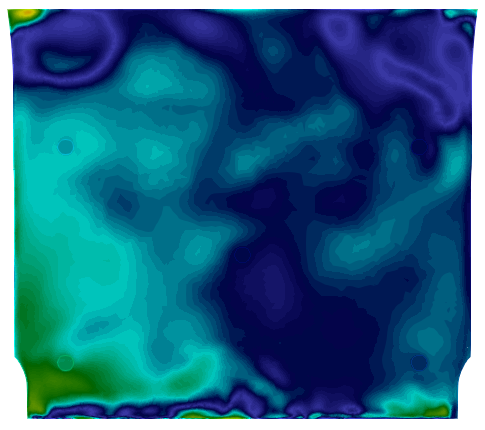}}
\subfloat[$z$ = 1.59975]{\label{fig:GPODerror11_5sensor}\includegraphics[width=.20\linewidth]{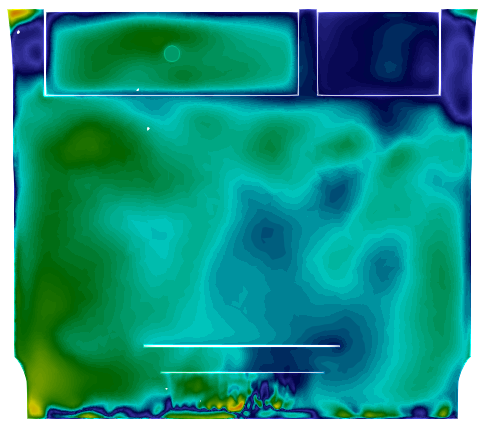}}
\end{subfigure}

\caption{Reconstruction error associated with the temperature field using the proposed ANNGPOD approach with 26 sensors.}
\label{fig:annGPOD_error}
\end{figure}

\fig{fig:GPODANNGPODerror_26sensors_para} shows the distribution of the absolute error in the middle plane, indicated as plane $11$ in \fig{fig:planes} while using sparse temperature data at $26$ sensor locations. The parametric location $p = 10$ is considered for this application. This analysis also demonstrates that the ANNGPOD always produces stable solutions whereas the conventional GPOD-based prediction can be completely erroneous as shown in \fig{fig:GPOD_MID} based on the number or location of the sensors chosen. We also show the absolute error distribution at all the planes corresponding to $p = 10$ and using $26$ sensor locations in \fig{fig:annGPOD_error}. However, the error distribution is comparable to the one demonstrated by the GPOD method using $5$ sensors and shown in \fig{fig:GPOD_error}, the potential of the ANNGPOD method lies in producing stable reconstruction at any number or locations of the sensors. Furthermore, the hyper-parameters of the ANN network can be tuned to improve the prediction accuracy which is not possible in the case of the conventional GPOD method. Additionally, the weights mentioned in Eqn. \ref{eq:anngpod}, can also be varied to improve the prediction accuracy i.e., $\lambda_1$ and $\lambda_2$ can be used to decide the amount of information or feedback necessary to be considered from the numerical datasets available at training parameters or from the numerical or experimental sparse dataset available at the sensor locations of validation parameters while predicting the full temperature field at all the spatial grid locations corresponding to the validation parameters' values. 

%As shown in \fig{fig:GPODANNGPOD_error_w1w21_var}, the $w1$ and $w2$ associated with the POD-ANN loss term and GPOD-based loss term respectively in Eqn. \ref{anngpod} is varied from $0$ to $1$ to assess their effect in the prediction accuracy corresponding to the parametric locations, $p = 10$ and $p = 62$. The $MAE$ error is greatly reduced using $w1$ of $0.1$ and $w1$ of $1$ which means providing more weights on the sparse dataset available at the sensor locations can improve the prediction accuracy. We also like to mention here that,  although the values of the weights may need to be varied to improve the prediction accuracy for other parameter values of interest, $w1$ and $w2$ work as additional control parameters in our proposed approach.      

\begin{figure}[ht]
\centering
\begin{subfigure}[b]{0.85\textwidth}
\centering
% \hspace{-3cm}
\subfloat[ $no.\ \ of \ \ sensors$ = 5, normal scale]{\label{fig:GPODANNGPOD_5sensorsn}\includegraphics[width=.50\linewidth]{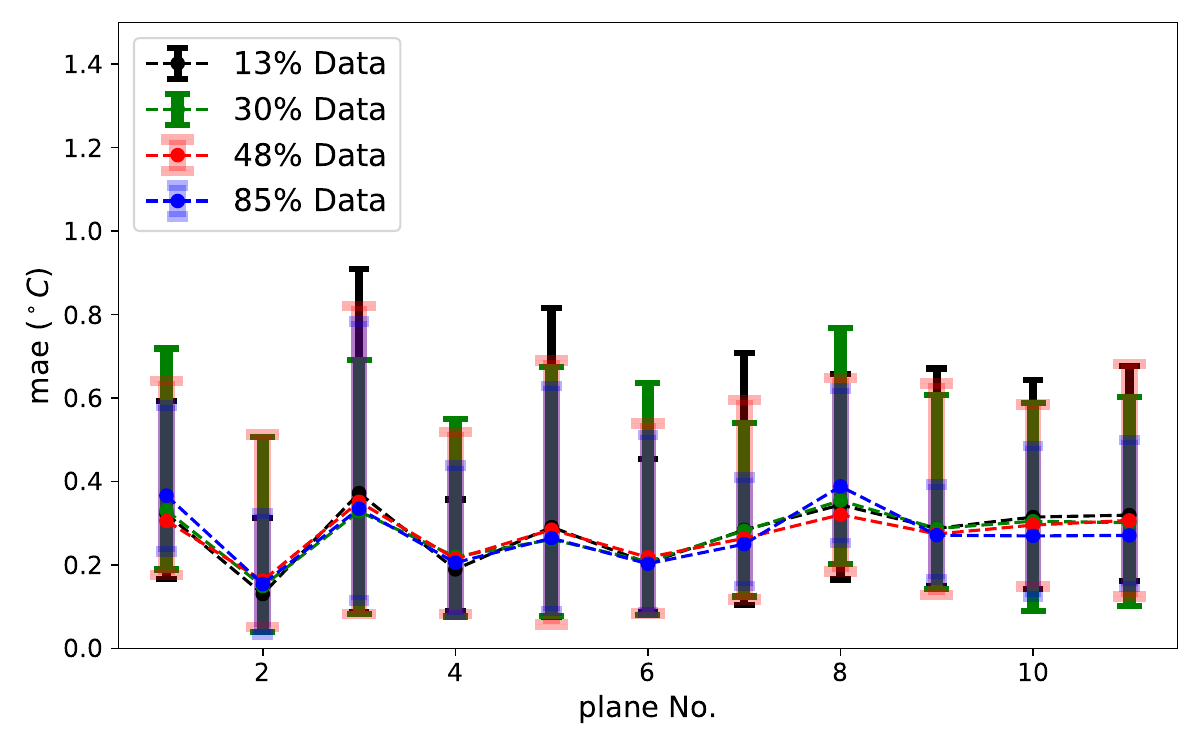}}
\subfloat[$no.\ \ of \ \ sensors$ = 5, log scale]{\label{fig:GPODANNGPOD_5sensorsl}\includegraphics[width=.50\linewidth]{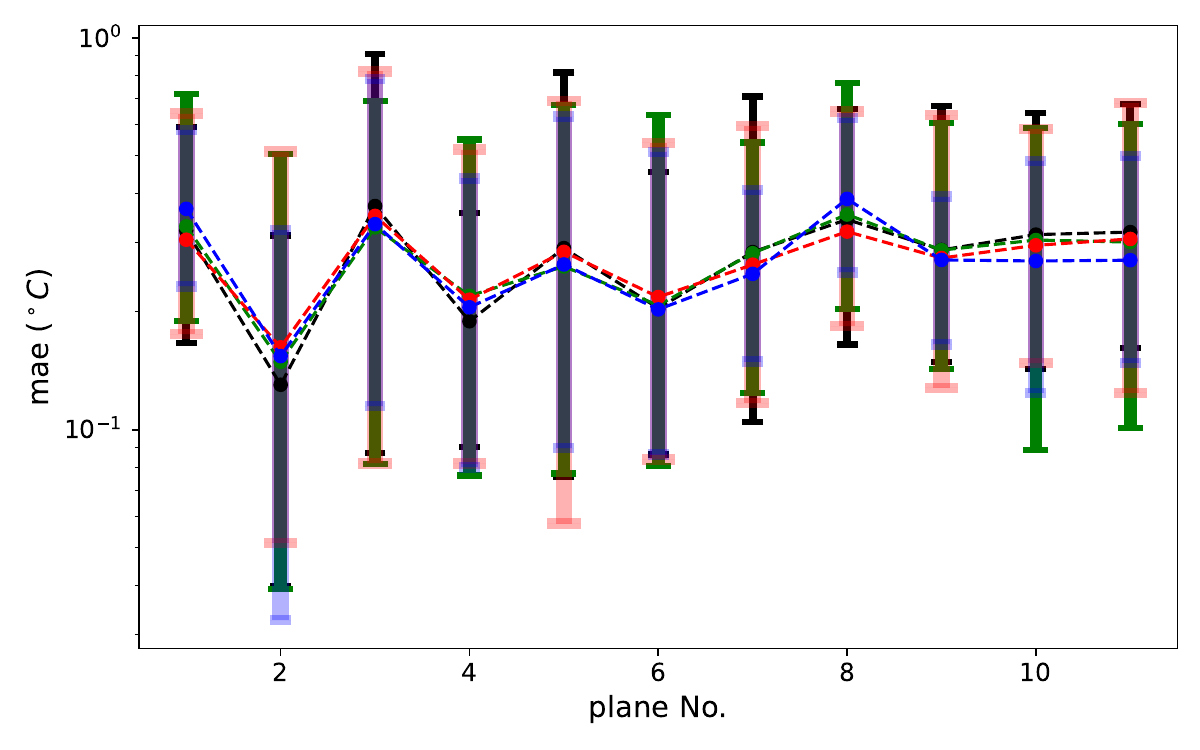}}
\end{subfigure}
\medskip
\begin{subfigure}[b]{0.85\textwidth}
\centering
% \hspace{-3cm}
\subfloat[ $no.\ \ of \ \ sensors$ = 26, normal scale]{\label{fig:GPODANNGPOD_26sensorsn}\includegraphics[width=.50\linewidth]{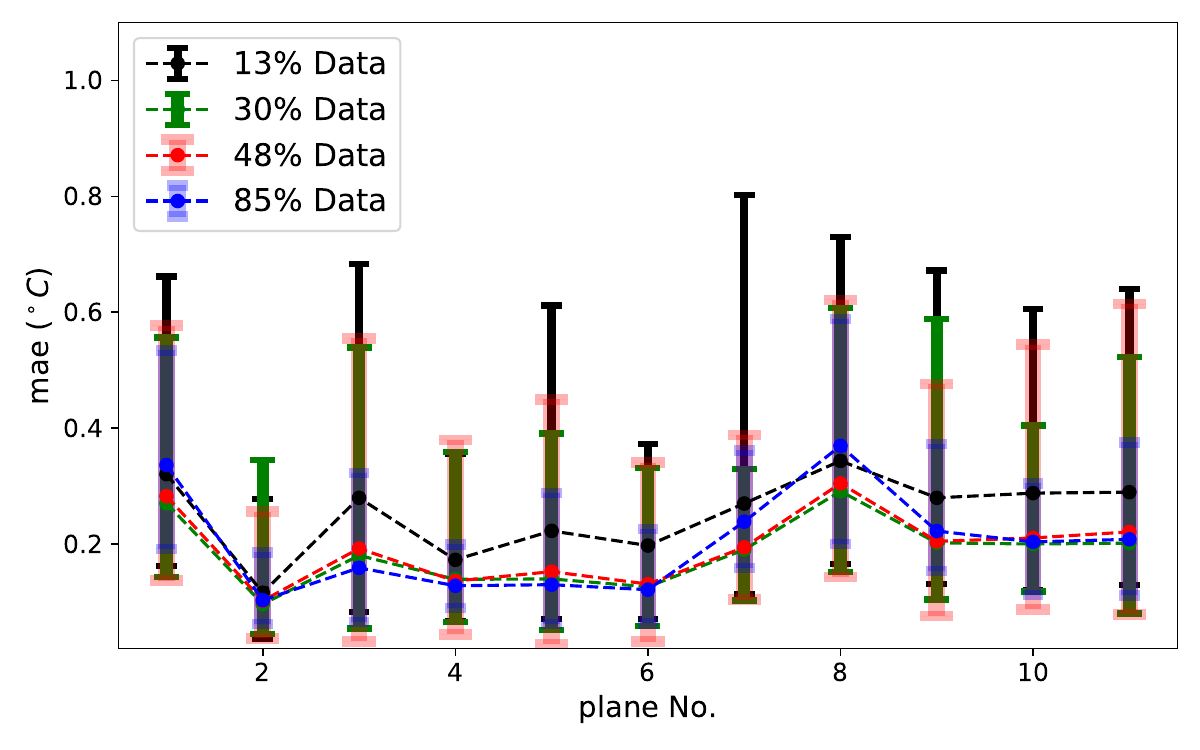}}
\subfloat[$no.\ \ of \ \ sensors$= 26, log scale]{\label{fig:GPODANNGPOD_26sensorsl}\includegraphics[width=.50\linewidth]{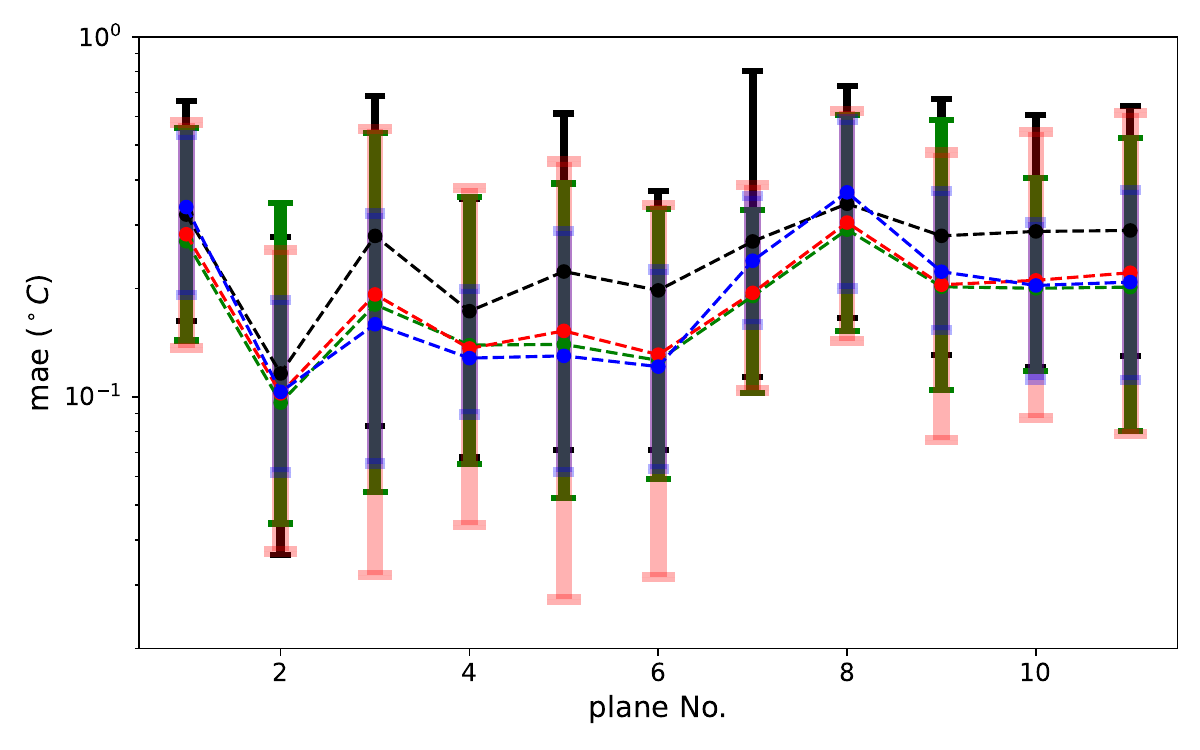}}
\end{subfigure}
\caption{Mean absolute error associated with the temperature field prediction using Gappy POD and ANNGPOD approach considering available data (numerical) at 5 and 26 sensor locations, respectively.}
\label{fig:GPODANNGPOD_sensors}
\end{figure}

Next, we consider sparse training datasets and compare the predicted results at all the validation points. It shows that with the usage of $5$ and $26$ sensors, the $\text{MAE}_\text{avg}$ and $\text{MAE}_\text{max}$ is less than $1$ $^o C$. For prediction using $26$ sensors and ANNGPOD method, $2$ inner layers with $200$ and $64$ neurons are considered. Furthermore, for the 5 sensors, $20$ neuron with one single layer is considered. With $85 \%$ case, the prediction error in terms of $\text{MAE}_\text{avg}$ and $\text{MAE}_\text{max}$ is minimum, and with sparsifying the training dataset, the prediction error increases. However, the prediction improvement is not monotonic with the increment in the training dataset, such as with the $48 \%$ data, the ANNGPOD method has shown poor performance than the $13 \%$ and $30 \%$ training data at most of the planes while using the number of sensors $5$ in \fig{fig:GPODANNGPOD_5sensorsl} and $26$ in \fig{fig:GPODANNGPOD_26sensorsl} respectively.  

% The error associated with different weights is as follows:
\begin{comment}
    
\begin{figure}[ht]
\centering
\begin{subfigure}[b]{0.85\textwidth}
\centering
% \hspace{-3cm}
\subfloat[ $p$ = 10, normal scale]{\label{fig:GPODANNGPOD_w11w21_10n}\includegraphics[width=.50\linewidth]{img/comparison_GPOD_ANNGPOD_w1w2_var/mae_comparison_GPOD_ANNGPOD_10_normalscale.pdf}}
\subfloat[$p$ = 10, log scale]{\label{fig:GPODANNGPOD_w11w21_10l}\includegraphics[width=.50\linewidth]{img/comparison_GPOD_ANNGPOD_w1w2_var/mae_comparison_GPOD_ANNGPOD_10_logscale.pdf}}
\end{subfigure}
\medskip
\begin{subfigure}[b]{0.85\textwidth}
\centering
% \hspace{-3cm}
\subfloat[ $p$ = 62, normal scale]{\label{fig:GPODANNGPOD_w11w21_62n}\includegraphics[width=.50\linewidth]{img/comparison_GPOD_ANNGPOD_w1w2_var/mae_comparison_GPOD_ANNGPOD_62_normalscale.pdf}}
\subfloat[$p$ = 62, log scale]{\label{fig:GPODANNGPOD_w11w21_62l}\includegraphics[width=.50\linewidth]{img/comparison_GPOD_ANNGPOD_w1w2_var/mae_comparison_GPOD_ANNGPOD_62_logscale.pdf}}
\end{subfigure}
\caption{Mean absolute error using Gappy POD and deep learning enhanced Gappy POD with a variation of w1 and w2}
\label{fig:GPODANNGPOD_error_w1w21_var}
\end{figure}
\end{comment}

\subsection{ANNGPOD with Experimental Sensor Data} \label{sec:res_anngpodexp} 
\begin{figure}[ht]
\centering
\begin{subfigure}[b]{0.85\textwidth}
\centering
% \hspace{-3cm}
\subfloat[Training Data = 98 $\%$]{\label{fig:exp98}\includegraphics[width=.50\linewidth]{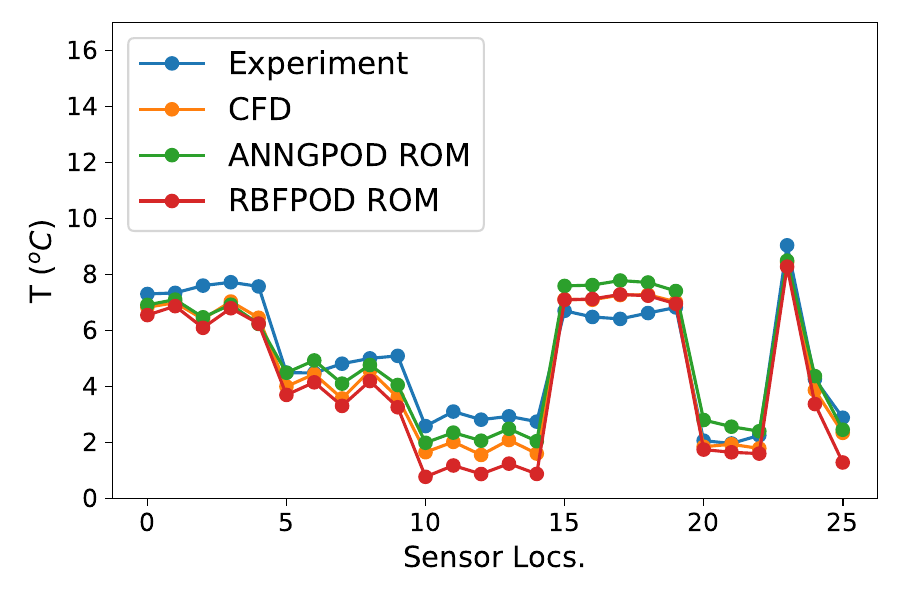}}
\subfloat[Training Data = 11 $\%$]{\label{fig:exp11}\includegraphics[width=.50\linewidth]{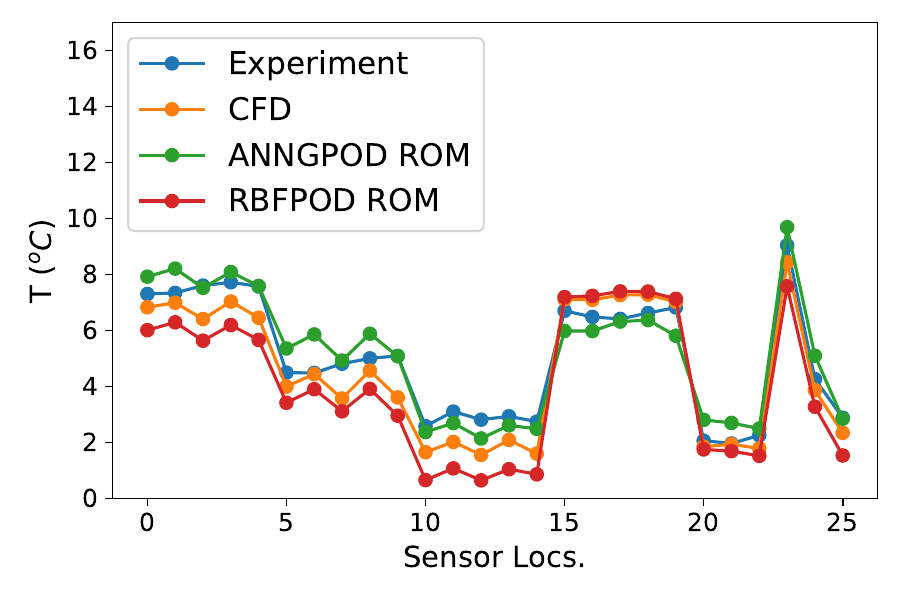}}
\end{subfigure}
\caption{Temperature prediction using the ANNGPOD method with available experimental data at 5 sensors' locations.}
\label{fig:Anngpod_Exp}
\end{figure}

We have considered the experimental result available for the parameter value $T_{ev}$ of $-15$ $^\circ C$, $T_{amb}$ of $32$ $^\circ C$, $v_f$ of $0 \%$. Out of 26 experimental sensors data available as shown in \fig{fig:sensor_loc}, we have considered only the middle sensors' datasets i.e., $1.3$, $2.3$, $3.3$, $tp1$ and $cr3$ as an input in ANNGPOD method. The temperature prediction is validated at 26 sensor locations where the experimental dataset is available. We have compared the prediction at the 26 points with CFD and the experiments when 13 $\%$ and 98 $\%$ training high-fidelity datasets are used respectively as shown in \fig{fig:Anngpod_Exp}. Table \ref{tab:exp_error} shows the mean absolute error of the predicted temperature using ANNGPOD and POD-RBF at the 26 sensor points with respect to the experiments and high-fidelity CFD results. As shown in Fig. \ref{fig:Anngpod_Exp} and Table \ref{tab:exp_error}, the POD-RBF ROM produces a mean absolute error of $0.3532$ and $0.5725$ using $98 \%$ and $13 \%$ CFD data when compared with the high-fidelity CFD simulation at the unknown parameter value. However, the prediction error with respect to experimental data at 26 sensor locations is significantly higher, $1.0273$ $^o C$ and $1.2562$ $^o C$. The POD-RBF-based prediction is completely based on the high-fidelity simulation data and does not receive any feedback from experimental results therefore, the surrogate model prediction moves away from the experimental results. As shown in \fig{fig:exp98} and \fig{fig:exp11}, the high-fidelity results are always below the experimental results at all the $26$ sensor locations. The RBF-POD ROM pushes the temperature prediction further down which results in a large mean absolute error with respect to the experiments. Conversely, in the ANNGPOD method, the prediction from the surrogate model takes $5$ sensors' temperature data as input from experiments and therefore reducing the mean absolute prediction error to $0.5925$ $^o C$ and $0.4960$ $^o C$ while using $98 \%$ and $13 \%$ numerical simulation data as training dataset. This is also important to note that when the $13 \%$ numerical simulation data is considered as a training dataset,  the prediction error of the surrogate model with respect to the experimental results is lower as compared to the one reported for the $98 \%$ training dataset. Since the ANNGPOD method is based on the minimization of loss terms derived from the numerical training dataset and experimental inputs, the later part (experimental) gets additional weightage when  $13 \%$ numerical CFD dataset is used.  

\begin{table}[h!]
\centering
 \begin{tabular}{||c |c | c||} 
 \hline
Average Error over 26 pts (in $K$) & POD-RBF  (98 $\%$) &  ANNGPOD (98 $\%$)  \\ [0.5ex] 
 \hline\hline
ROM vs CFD & 0.3532  & 0.2310  \\ %[1ex] 
Exp vs CFD & 0.6837  & 0.6837  \\ %[1ex]
ROM vs Exp & 1.0273  & 0.5925  \\ %[1ex]
 \hline
 \end{tabular}
 \begin{tabular}{||c |c | c||} 
 \hline
Average Error over 26 pts (in $K$) & POD-RBF  (11 $\%$) &  ANNGPOD (11 $\%$)  \\ [0.5ex] 
 \hline\hline
ROM vs CFD & 0.5725  & 1.0253  \\ %[1ex] 
Exp vs CFD & 0.6837  & 0.6837  \\ %[1ex]
ROM vs Exp & 1.2562  & 0.4960  \\ %[1ex]
 \hline
 \end{tabular}
 \caption{prediction accuracy of ANNGPOD method with respect to the experimental results at 5 mid sensors}
 \label{tab:exp_error}
\end{table}

\subsection{Computational cost}\label{sec:compeff}

We briefly discuss the efficiency of our ROM approach. 
The FOM simulations were carried out on a supercomputer using 128 processes. The computing nodes  were equipped with Xeon E5-2683 v4 processors (2 sockets, 32 cores) and 64GB RAM. The CPU time associated with the FOM simulation is around $1e6$ s. The ROM simulations were conducted on an 11th Gen Intel(R) Core(TM) i7-11700 @ 2.50GHz 32GB RAM by using one only processor. The training time associated with the POD-RBF ROM is $35.24$ $s$ while the prediction time at all the 11 planes of interest is $0.64$ $s$. Conversely, in the case of ANNGPOD ROM, there are no distinctions between the training and prediction phases, for each unknown parameter value we need to rerun the algorithm for the reconstruction of the temperature field. Therefore the approximate time taken for the ANNGPOD prediction is $193.01$ $s$. It is evident from the computational cost associated with the POD-RBF and ANNGPOD ROM that, the POD-RBF ROM can be used for the real-time analysis while the latter can not be used which is the limitation of our present approach. However, a significant computational speed-up is achieved, i.e., for POD-RBF ROM, it is $1e6$ and for ANNGPOD it is $5e3$ times.    

%\newpage \clearpage
%\clearpage
\section{Concluding remarks}\label{sec:conc}

In This study, the thermal behaviour of a household fridge was investigated through Conjugate Heat Transfer (CHT) simulation. To ensure the accuracy and reliability of the suggested model, it was validated against experimental setup for two various configurations: static and ventilated fridges. Moreover, we conducted a sensitivity analysis for the key parameters of this problem. The result identified that creating a tailored database for this problem requires denser sampling of the fan velocity, as the most sensitive parameter. This database forms the foundation for the development of a non-intrusive data-driven Reduced Order Model (ROM) for the fridge. In this work, we propose two different types of ROM: POD-RBF and ANNGPOD methods for the exploration of the parametric space - the former considers only high-fidelity numerical data as a training dataset while the latter can use both experimental and numerical data for the temperature field prediction at $11$ planes of interest. We have shown that the POD-RBF ROM  produces a mean absolute error of less than $1$ $^o C$ with respect to the high-fidelity simulation results at all $11$ planes while using a very sparse training dataset of $11$ $\%$. We have proposed a novel parametric ROM ANNGPOD capable of taking both experimental and numerical data as inputs. It outperforms the conventional Gappy POD approach producing a very accurate prediction error below the limit of $1$ $K$ at very sparse datasets and at any number of sensors while the classical Gappy POD produces accurate prediction only when a selected number of sensors are considered using a relatively richer training dataset. While carrying out this comparison we have solely considered the numerical datasets for both methods. Finally, the experimental datasets are considered at $5$ sensor locations as input alongside the numerical training dataset at selected training parameter values and we have demonstrated that the ANNGPOD produced very accurate predictions close to the experiments at 26 sensor locations as compared to the POD-RBF ROM which does not take any feedback from the experiments and thereby demonstrating the potential of our novel approach, ANNGPOD. Looking ahead, recent advances in hybrid deep-learning architectures, such as  \cite{hajisharifi2025combining}, highlight promising directions for further enhancing the predictive capability of ROMs in CHT problems.

\section{acknowledgments}\label{sec:akw}

We acknowledge the support provided by the European Research Council Executive Agency by the Consolidator Grant project AROMA-CFD "Advanced Reduced Order Methods with Applications in Computational Fluid Dynamics" - GA 681447, H2020-ERC CoG 2015 AROMA-CFD, the European Union’s Horizon 2020 research and innovation program under the Marie Skłodowska-Curie Actions, grant agreement 872442 (ARIA), PON “Research and Innovation on Green related issues” FSE REACT-EU 2021 project, PRIN NA FROM-PDEs project, INdAM-GNCS 2019-2021 projects. We also would like to  acknowledge Electrolux Company for their valuable support and fruitful discussion. Their expertise and assistance greatly contributed to the success of this research.

\clearpage
\bibliographystyle{plain}
\bibliography{mybib}

\end{document}